\documentclass[fleqn,usenatbib,useAMS]{mnras}

\usepackage{mathptmx}

\usepackage[T1]{fontenc}
\usepackage{ae,aecompl}

\usepackage{color}

\usepackage{graphicx}	
\usepackage{amsmath}	
\usepackage{amssymb}	

\newcommand{\hb}{H$\beta$}
\newcommand{\hi}{H\thinspace\textsc{i}}
\newcommand{\foiii}{[O\thinspace\textsc{iii}]}
\newcommand{\foii}{[O\thinspace\textsc{ii}]}
\newcommand{\fsii}{[S\thinspace\textsc{ii}]}
\newcommand{\fsiii}{[S\thinspace\textsc{iii}]}
\newcommand{\fnii}{[N\thinspace\textsc{ii}]}
\newcommand{\fcliii}{[Cl\thinspace\textsc{iii}]}
\newcommand{\cii}{C\thinspace\textsc{ii}}
\newcommand{\oii}{O\thinspace\textsc{ii}}
\newcommand{\hii}{H\thinspace\textsc{ii}}
\newcommand{\sfciii}{C\thinspace\textsc{iii}]}
\newcommand{\sfcii}{C\thinspace\textsc{ii}]}
\newcommand{\hei}{He\thinspace\textsc{i}}
\newcommand{\heii}{He\thinspace\textsc{ii}}

\title[C and O in the Magellanic Clouds]{Carbon and oxygen in H\,{\Large \textbf{II}} regions of the Magellanic Clouds: abundance discrepancy and chemical evolution}

\author[L.~Toribio San Cipriano et al.]
{L.~Toribio San Cipriano,$^{1,2}$\thanks{E-mail: ltoribio@iac.es}
G.~Dom\'inguez-Guzm\'an,$^{3}$
C.~Esteban,$^{1,2}$\newauthor
J.~Garc\'ia-Rojas,$^{1,2}$
A.~Mesa-Delgado,$^{4}$
F.~Bresolin,$^{5}$
M.~Rodr\'iguez,$^{3}$\newauthor
and S.~Sim\'on-D\'iaz$^{1,2}$
\\
$^{1}$Instituto de Astrof\'isica de Canarias, E-38200, La Laguna, Tenerife, Spain\\
$^{2}$Departamento de Astrof\'isica, Universidad de La Laguna, E-38206, La Laguna, Tenerife, Spain\\
$^{3}$Instituto Nacional de Astrof\'isica, \'Optica y Electr\'onica, Apartado Postal 51 y 216, Puebla, Mexico\\
$^{4}$Instituto de Astrof\'isica, Facultad de F\'isica, Pontificia Universidad Cat\'olica de Chile, Av. Vicu\~na Mackenna 4860,\\ 
782-0436 Macul, Santiago, Chile \\
$^{5}$Institute for Astronomy, 2680 Woodlawn Drive, Honolulu, HI 96822, USA
}

\date{Accepted 2017 February 3. Received 2017 February 3; in original form 2016 November 25.}

\pubyear{2017}

\begin{document}
\label{firstpage}
\pagerange{\pageref{firstpage}--\pageref{lastpage}}
\maketitle

\begin{abstract}

We present C and O abundances in the Magellanic Clouds derived from deep spectra of \hii\ regions. The data have been taken with the Ultraviolet-Visual Echelle Spectrograph at the 8.2-m VLT. The sample comprises 5 \hii\ regions in the Large Magellanic Cloud (LMC) and 4 in the Small Magellanic Cloud (SMC). We measure pure recombination lines (RLs) of \cii\ and \oii\ in all the objects, permitting to derive the abundance discrepancy factors (ADFs) for O$^{2+}$, as well as their O/H, C/H and C/O ratios. We compare the ADFs with those of other \hii\ regions in different galaxies. The results suggest a possible metallicity dependence of the ADF for the low-metallicity objects; but more uncertain for high-metallicity objects. We compare nebular and B-type stellar abundances and we find that the stellar abundances agree better with the nebular ones derived from collisionally excited lines (CELs). Comparing these results with other galaxies we observe that stellar abundances seem to agree better with the nebular ones derived from CELs in low-metallicity environments and from RLs in high-metallicity environments. The C/H, O/H and C/O ratios show almost flat radial gradients, in contrast with the spiral galaxies where such  gradients are negative. We explore the chemical evolution analysing C/O vs. O/H and comparing with the results of \hii\ regions in other galaxies. The LMC seems to show a similar chemical evolution to the external zones of small spiral galaxies and the SMC behaves as a typical star-forming dwarf galaxy.

\end{abstract}




\section{Introduction}

The knowledge of the chemical composition of extragalactic \hii\ regions is crucial for building models that reproduce the chemical evolution of galaxies. In particular, O/H, C/H and C/O ratios are used as tracers of the chemical enrichment of the discs of galaxies. Carbon (C) is the second most abundant heavy element, after oxygen, in the Universe, is an important source of opacity in stars and one of the main elements found in interstellar dust and organic molecules. In addition, C plays an important role to constrain chemical evolution models of galaxies \citep[e.g.][]{2005ApJ...623..213C}. Despite its significance, C abundances in extragalactic \hii\ regions have been poorly explored. \citet{1995ApJ...443...64G, 1999ApJ...513..168G} derived C abundances from HST/FOS spectrophotometry in several extragalactic \hii\ regions of nearby spiral galaxies. They derived the abundances using the \sfciii\ 1909 \AA\ and \sfcii\ 2326 \AA\ collisionally excited lines (hereinafter CELs), which are strongly affected by the uncertainty in the choice of UV reddening function. More recently, \citet{2016ApJ...827..126B} reported C abundances also determined from CELs for 12 \hii\ regions in 12 nearby dwarf galaxies, obtained through HST/COS spectrophotometry. On the other hand, there is an alternative method to derive C abundances in ionized nebulae, which is based on the measurement of the faint \cii\ 4267 \AA\ recombination line (hereinafter RL). In the last years, this line has been accessible in Galactic and extragalactic \hii\ regions thanks to the arrival of large aperture ground-based optical telescopes \citep[e.g.][]{1986A&A...158..266P, 2005ApJ...634.1056P, 2002ApJ...581..241E, 2009ApJ...700..654E, 2014MNRAS.443..624E, 2007ApJ...656..168L, 2016MNRAS.458.1866T}.

Traditionally, CELs have been used to derive the chemical abundances in \hii\ regions because they are brighter and easier to detect than RLs. However, the abundances derived from RLs are much less sensitive to the adopted electron temperature and density than the CELs abundances. A well-known and unsolved issue in the astrophysics of ionized nebulae is the so-called abundance discrepancy problem, which consists of the fact that chemical abundances are different depending on the type of lines (CELs or RLs) used to derive them. The abundances derived from RLs are systematically higher than those derived from CELs. Several authors have proposed different hypotheses to explain this problem \citep{1969BOTT....5....3P, 2005MNRAS.364..687T, 2012ApJ...752..148N} but the debate is still open. \citet{2007ApJ...670..457G} analyzed the behaviour of the abundance discrepancy in several Galactic and extragalactic \hii\ regions. These authors reported that the abundance discrepancy factor (ADF)\footnote{
Defined as: 
\begin{equation}
\mathrm{ADF(X^{i+}) \equiv log(X^{i+}/H^+)_{RLs} - log(X^{i+}/H^+)_{CELs};}
\end{equation}
where $\mathrm{X^{i+}}$ corresponds to the ionization state i of element X.} for O$^{++}$ is fairly constant and of the order of 2 in the studied sample. On the other hand, \citet{2011BSRSL..80..255S} compared the Orion nebula abundances with those of its associated OB-type stars -- which should have the same chemical composition -- in order to clarify which type of emission lines provides the `true' abundances. They found that oxygen (O) abundances from RLs agree better with the stellar O ones than the abundances from CELs.

The Large Magellanic Cloud (LMC) and the Small Magellanic Cloud (SMC) form an interacting system of galaxies of the Local Group. Both objects are known as the Magellanic Clouds (MCs) and are satellites of the Milky Way. The first studies about chemical abundances in the MCs determined using spectra of \hii\ regions were carried out by \citet[][4 objects in LMC and 3 in SMC]{1974ApJ...193..327P, 1976ApJ...203..581P} and \citet[][11 in LMC and 3 in SMC]{1975ApJ...195..315D}. \citet{1978MNRAS.184..569P} extended the sample with 6 \hii\ regions in the LMC and 17 in the SMC and discussed the spatial distribution of O abundances in both galaxies. Additional studies of the chemical composition of \hii\ regions in the MCs are those of \citet[][NGC\,346 in the SMC]{2000ApJ...541..688P} and \citet[][N77 in the LMC]{2012A&A...545A..29S}. Only \citet[][N88A in the SMC and 30 Doradus in the LMC]{1995ApJ...443...64G} and \citet[][3 \hii\ regions in the SMC and 4 in the LMC]{1982ApJ...252..461D} reported C and O abundances of \hii\ regions derived from CELs in the UV.  \citet[][N66 in the SMC and 30 Doradus and N11B in the LMC]{2003MNRAS.338..687T}, \citet[][30 Doradus in the LMC]{2003ApJ...584..735P} and \citet[][NGC\,456 in the SMC]{2012ApJ...746..115P} determined O and C abundances using optical RLs.

In this work we present new results based on the deepest and most comprehensive collection of optical spectra of \hii\ regions in the MCs. This set of data will be described in depth and further analysed in Dom\'inguez-Guzm\'an et al. (in preparation). 
The high quality of the data allows us to measure the faint \cii\ 4267 \AA\ RL and all or most of the lines of multiplet 1 of the \oii\ RLs at about 4650 \AA, permitting the calculation of $\mathrm{{C}^{2+}}$ and $\mathrm{{O}^{2+}}$ abundances. These determinations allow us to explore the abundance discrepancy problem, as well as to analyze the spatial distribution of O/H, C/H and C/O ratios across the MCs and to compare the results with those of other galaxies. 

This paper is organized as follows. In Section~\ref{sec:sample} we describe the data acquisition and reduction procedures. In Section~\ref{sec:lines} 
we describe the line measurement process and the reddening correction. In Section~\ref{sec:results} we compute the physical conditions and the chemical abundances of the ionized gas. In Section~\ref{sec:discuss} we discuss the abundance discrepancy problem and we compare the nebular and stellar abundances. In addition, we analyze the spatial distribution of O/H, C/H and C/O ratios across the galaxies. We summarize our results in Section~\ref{sec:conclusion}.

\section{Sample and data reduction}\label{sec:sample}

\begin{figure}
 \centering  
  \includegraphics[width=0.48\textwidth]{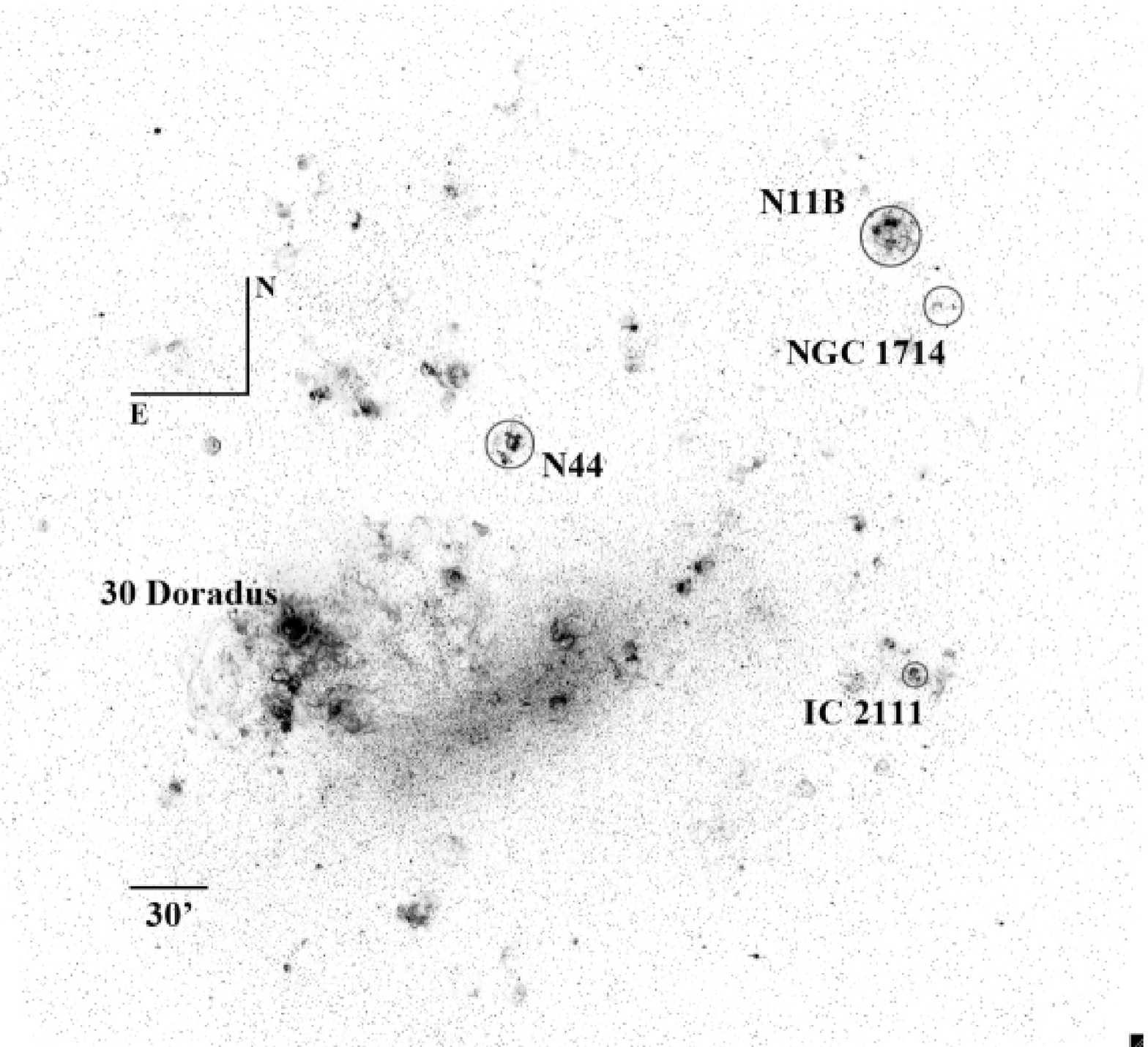}
  \includegraphics[width=0.48\textwidth]{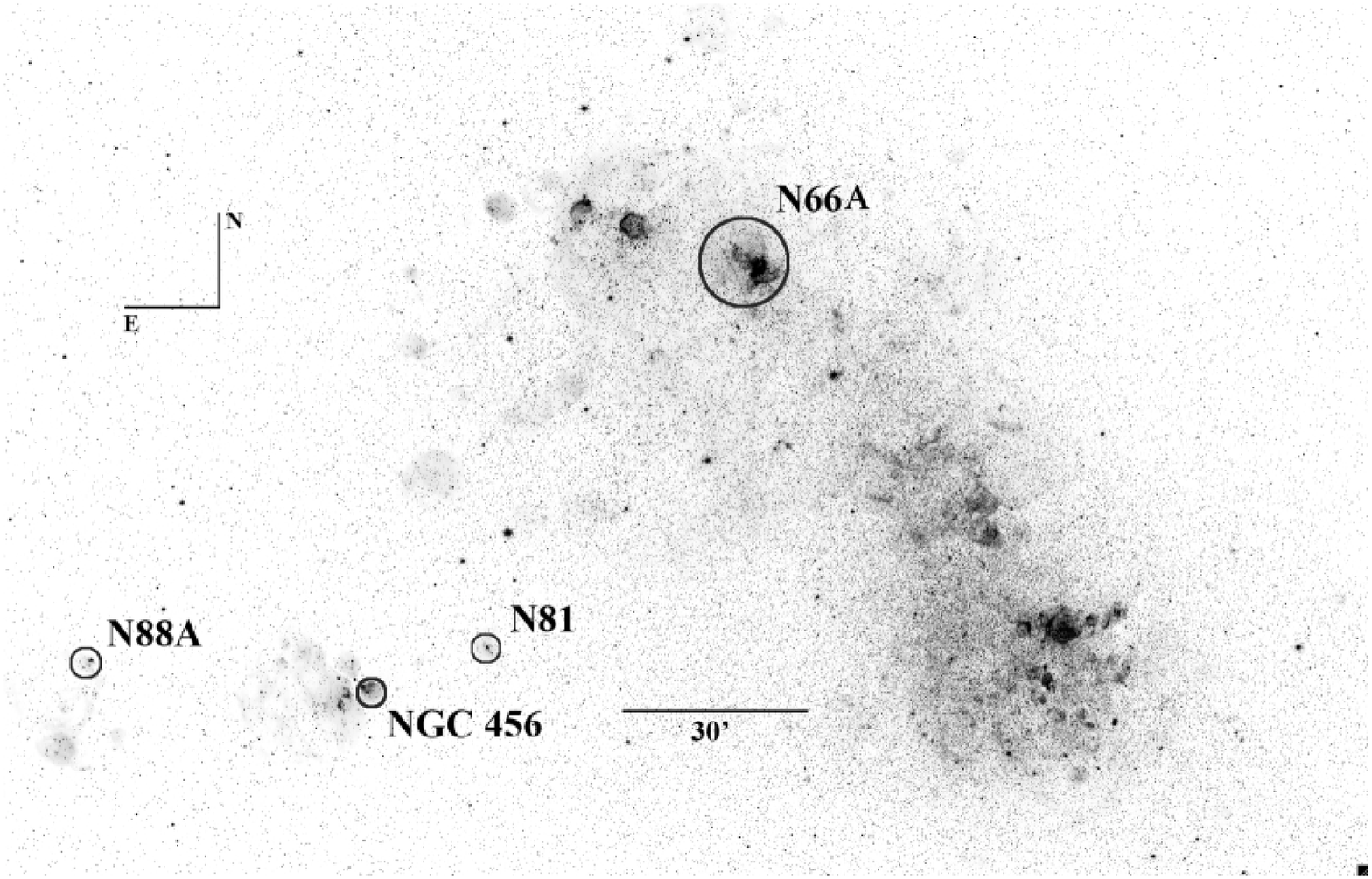}
 \caption{Distribution of the observed \hii\ regions in the LMC (upper panel) and in the SMC (lower panel). Credits: \citet{1988AJ.....96..877B, 1995AJ....109..594K, 1998AJ....116..180P}.}
 \label{fig:mapa}
\end{figure}

The sample considered in this work comprises 9 \hii\ regions: 5 in the LMC and 4 in the SMC. The distribution of the \hii\ regions in each galaxy is shown in Fig.~\ref{fig:mapa}. We observed 7 of these objects, while the data for 30 Doradus and NGC\,456 were taken from \citet{2003ApJ...584..735P} and \citet{2012ApJ...746..115P}. However, to have an homogeneous data set, in these two cases, we took the measured line fluxes and performed the analysis in the same way than for the rest of the sample.

The observations were made on 2003 March and on 2013 November at Cerro Paranal Observatory (Chile) with the Ultraviolet Visual Echelle Spectrograph (UVES) mounted at the Kueyen unit of the 8.2-m Very Large Telescope (VLT). We used the standard settings: DIC1(346+580) and DIC2(437+860) covering the spectral range 3100--10420 \AA. The wavelength intervals 5782--5824 and 8543--8634 \AA\ were not observed due to a gap between the two CCDs used in the red arm. Additionally, there are also two small spectral gaps between 10083--10090 and 10251--10263 \AA\ because the redmost orders did not fit completely within the CCD. The atmospheric dispersion corrector was used to keep the same observed region within the slit regardless of the airmass value. The objects were observed at airmass values between 1.35 and 1.88. The slit width was set to 2 arcsec for the NGC\,1714 region and 3 arcsec for the rest of the regions, while the slit length was 10 arcsec in the red arm and 12 arcsec in the blue one. The slit width was chosen in order to maximize the signal-to-noise ratio (S/N) and to have enough spectral resolution ($R \sim 8800$) to separate the relevant faint lines for this study. The position angle of the slit and the area extracted for each \hii\ region were selected to cover most of their core extension. 

We summarize some of the main parameters of the LMC and the SMC in Table~\ref{tab:parameters}. The properties and description of the observations of the sample of \hii\ regions are shown in Table~\ref{tab:information}.

\begin{table}
   \centering
   \caption{Main parameters of the LMC and SMC.}
   \label{tab:parameters}
   \begin{tabular}{lcccc} 
        \hline
        Parameter             & LMC         & Ref.   & SMC        & Ref.\\
        \hline                                           
         RA  (J2000)          &  05:25:06    & 1   &  00:52:44    &  3 \\
         Dec (J2000)          & -69:47:00    & 1   & -72:49:43    &  3 \\
         Distance (kpc)       &  49.97       & 1   &  61.94       &  3 \\
         $R_{25}$ (kpc)       &  4.74        & 2   &  2.85        &  2 \\
         Inclination ($\degr$)&  28          & 1   &  68          &  4  \\
         PA ($\degr$)         &  12          & 1   &  238         &  4   \\
         $M_{\mathrm V}$      &  -18.35      & 2   &  -17.04      &  2    \\         
        \hline
        \multicolumn{5}{l}{1.- \citet{2013Natur.495...76P}.}\\ 
        \multicolumn{5}{l}{2.- Calculated using $D_{25}$ by \citet{1991S&T....82Q.621D}}.\\
        \multicolumn{5}{l}{3.- \citet{2016ApJ...816...49S}.}\\
        \multicolumn{5}{l}{4.- \citet{2000A&A...363..901G}.}\\
   \end{tabular}
\end{table}

\begin{table*}
   \centering
   \caption{Properties and description of the observations of \hii\ regions in the LMC and SMC.}
   \label{tab:information}
    \begin{tabular}{l @{\hspace{0.1cm}}cccccccccc@{}|p{0.8cm}|c}
        \hline
              &                      & R.A.        & Decl.        & $R_\mathrm{G}^{c}$ &        & \multicolumn{2}{c}{Exposure time (s)}               & Extracted area         &    PA       &Additional \\
        Cloud &  Object              & (J2000)     & (J2000)      &    (kpc)       &$R / R_{25}^{d}$&  DIC1(346+580)               &   DIC2(437+860)              & ($\mathrm{arcsec}^2$)  &    ($\degr$)&   Info     \\
        \hline                                                                                                                                                                  
         LMC  &30 Doradus$^\mathrm{a}$& 05:38:42.3 & -69:06:03.0   &    1.29       & 0.30       &        --                    &        --                    &        --              &   --        &                \\
         LMC  & N44C$^\mathrm{b}$    & 05:22:13.6  & -67:58:34.2   &    1.62       & 0.37       &$3 \times 30$, $3 \times 300$ &$3 \times 30$, $3 \times 1200$&$3.0 \times 9.4$        &   90        &                    \\
         LMC  & IC\,2111             & 04:51:52.1  & -69:23:32.0   &    2.88       & 0.66       &$1 \times 60$, $3 \times 240$ &$1 \times 60$, $3 \times 700$ &$3.0 \times 9.4$        &   90        & N79A$^\mathrm{b}$ \\
         LMC  & NGC\,1714            & 04:52:08.8  & -66:55:24.0   &    3.99       & 0.91       &$3 \times 300$                &$3 \times 900$                &$2.0 \times 9.4$        &  315        &                     \\
         LMC  & N11B$^\mathrm{b}$    & 04:56:46.9  & -66:24:37.9   &    4.03       & 0.93       &$3 \times 30$, $4 \times 900$ &$3 \times 30$, $4 \times 2000$&$3.0 \times 9.4$        &  120        &          NGC\,1763  \\
         SMC  & N66A$^\mathrm{b}$    & 00:59:14.3  & -72:11:02.8   &    1.16       & 0.42       &$3 \times 30$, $3 \times 600$ &$3 \times 30$, $3 \times 1500$&$3.0 \times 9.4$        &    0        &         NGC\,346      \\
         SMC  & N81$^\mathrm{b}$     & 01:09:12.8  & -73:11:36.9   &    2.85       & 1.03       &$3 \times 30$, $3 \times 600$ &$3 \times 30$, $3 \times 1800$&$3.0 \times 9.4$        &   90        &               \\
         SMC  & NGC\,456$^\mathrm{a}$& 01:13:44.4  & -73:17:26.0   &    3.62       & 1.31       &         --                   &         --                   &       --               &   --        &    N83A$^\mathrm{b}$  \\
         SMC  & N88A$^\mathrm{b}$    & 01:24:08.3  & -73:09:04.6   &    4.67       & 1.69       &$3 \times 30$, $3 \times 800$ &$3 \times 30$, $3 \times 1200$&$3.0 \times 5.3$        &   110       &               \\
        \hline
        \multicolumn{10}{l}{$^\mathrm{a}$ Data taken from the literature: \citet[][30 Doradus]{2003ApJ...584..735P} and \citet[][NGC\,456]{2012ApJ...746..115P}.}\\
        \multicolumn{10}{l}{$^\mathrm{b}$ Target names following the catalogue of \citet{1956ApJS....2..315H}.}\\
        \multicolumn{10}{l}{$^\mathrm{c}$ Deprojected Galactocentric distance computed using information of Table~\ref{tab:parameters}.}\\
        \multicolumn{10}{l}{$^\mathrm{d}$ Deprojected Galactocentric distance normalized to the isophotal radius ($R_{25}$) given in Table~\ref{tab:parameters}.}\\

   \end{tabular}
\end{table*}

The spectra were reduced using the public ESO UVES pipeline \citep{2000Msngr.101...31B} under the \textsc{gasgano} graphic user interface, following the standard procedure of bias subtraction, order tracing, aperture extraction, background subtraction, flat-fielding and wavelength calibration. The final products of the pipeline are 2D wavelength calibrated spectra. The spectrophotometric standard stars HR 9087, HR 718 and HR 3454 were observed to perform the flux calibration. We estimated a flux calibration error $\sim 4$ per cent for the \hii\ regions observed in the  2003 campaign (IC\,2111 and NGC\,1714) and $\sim 1 - 2.5$ per cent for those observed in 2013. We used \textsc{iraf}\footnote{\textsc{iraf} is distributed by National Optical Astronomy Observatory, which is operated by Association of Universities for Research in Astronomy, under cooperative agreement with the National Science Foundation.} to obtain the final one-dimensional flux calibrated spectra.   

\section{Line intensity measurement and reddening correction}\label{sec:lines}

\begin{figure}
 \centering
  \includegraphics[]{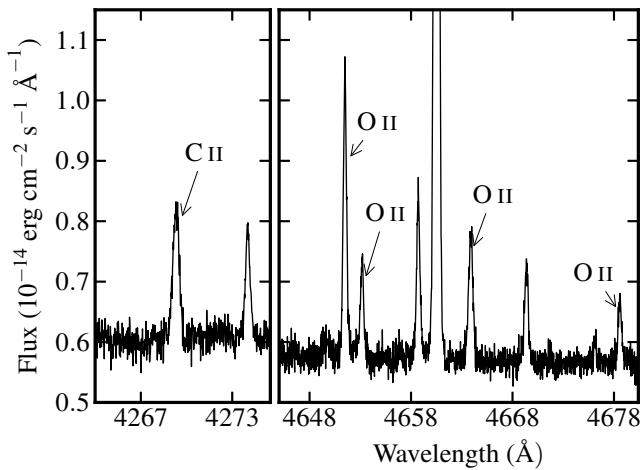}
 \caption{Example of a section of the UVES spectra of the \hii\ region N88A of the SMC, which shows the RL of \cii\ 4267 \AA\ (left-hand panel) and the RLs of multiplet 1 of \oii\ around 4650 \AA\ (right-hand panel).}
 \label{fig:example_spec}
\end{figure}

We measured the fluxes of the lines needed to derive the reddening coefficient -- Balmer and Paschen {\hi} lines --, the physical conditions of the ionized gas, as well as the \cii\ and \oii\ RLs for deriving the $\mathrm{C}^{2+}$ and $\mathrm{O}^{2+}$ abundances, which are the main goals of this work. In addition, in the case of N44C in the LMC, we also measured several {\hei} and {\heii} lines because we detect {\heii} 4686 \AA\ emission in its spectrum. In Fig.~\ref{fig:example_spec} we show the \cii\ 4267 \AA\ RL and those of multiplet 1 of \oii\ 4650 \AA\ of our flux-calibrated VLT UVES spectrum of the \hii\ region N88A in the SMC. The spectra of the objects show a much larger number of emission lines, but their line intensities and analysis will be presented in Dom\'inguez-Guzm\'an et al. (in preparation).  

Line intensities were measured integrating all the flux in the line between two given limits and over a local continuum estimated by eye. In the case of line blending we fitted a double Gaussian profile to measure the individual line fluxes. All the measurements were made with the \textsc{splot} task of \textsc{iraf}. 

The line intensities were normalized to \hb. The DIC2\,437 and DIC1\,580 configurations contain the \hb\ line, but for the bluest spectrum (DIC1\,346 configuration), we normalized to the H9 line and then re-scaled to \hb\ using the H9/\hb\ ratio measured in DIC2\,437.  For the reddest spectrum (DIC2\,860 configuration),  we normalized to \fsii\ 6716 \AA\ line and then we re-scaled to \hb\ using the \fsii\ 6716/\hb\ ratio observed in DIC1\,580.
 
We corrected the observed line intensity ratios for the effect of interstellar reddening. We assumed that the equivalent width of the absorption components due to underlying stars is the same for all the \hi\ lines. We determined the reddening coefficient, $c$({\hb}), following the equation from \citet{2006A&A...449..997L}: 

\begin{equation}
c(\mathrm{H}\beta) = \frac{1}{f(\lambda)}\log \left[\frac{\frac{I(\lambda)}{I(\mathrm{H}\beta)}  \left(1 + \frac{W_{\mathrm{abs}}}{W_{\mathrm{H}\beta}}\right)}{\frac{F(\lambda)}{F(\mathrm{H}\beta)} \left(1 + \frac{W_{\mathrm{abs}}}{W_{\lambda}}\right)}\right] ,\label{eq: c(Hbeta)}
\end{equation}
where $f(\lambda)$ is the reddening function, $F(\lambda)$/$F$({\hb}) is the observed line intensity ratio with respect to \hb, $I(\lambda)$/$I$({\hb}) is the theoretical ratio, $W_{\mathrm{abs}}$,  $W_{\lambda}$ and $W_{\rm H\beta}$ are the equivalent widths of the underlying stellar absorption, the considered \hi\ line and H$\beta$, respectively. We computed $c$({\hb}) using the Balmer lines from H15 to H3, excluding H14, H10 and H8 because they are blended with other lines, and the Paschen lines from P20 to P7. The theoretical line ratios were obtained from \citet{1995MNRAS.272...41S} for case B assuming $n_\mathrm{e} = 100\ \mathrm{cm^3}$ and $T_\mathrm{e} = 10000\ \mathrm{K}$ and the reddening function by \citet{1983MNRAS.203..301H}.  We performed an iterative process to derive $c$({\hb}) with different $W_{\mathrm{abs}}$ and choosing the $c$({\hb}) that best fits the observed-to-theoretical line ratios. We finally obtained the best fit for all the objects when $W_{\mathrm{abs}} = 0$. The value of $c$({\hb}) adopted for each \hii\ region is shown in Tables~\ref{tab:lines_lmc} and \ref{tab:lines_smc} of appendix \ref{appex:1}. 

Tables~\ref{tab:lines_lmc} and \ref{tab:lines_smc} of appendix \ref{appex:1} list the laboratory wavelength, $f(\lambda)$, and the reddening corrected intensity ratio with respect to \hb\ along with its associated uncertainty for each of the lines used in this paper. We calculated the uncertainty of each intensity ratio by adding quadratically the uncertainty in the measured line flux, the uncertainty associated to the reddening coefficient, and the uncertainty of the flux calibration. For 30 Doradus and NGC\,456 we computed the uncertainties considering the errors in the line intensity ratios given in the original references. The uncertainties of the measured line fluxes were determined following the equation from \citet{1999MNRAS.310..262T}:
\begin{equation}
\sigma_F = \sigma_c D \sqrt{2N_{pix} + EW / D},
\end{equation}
where $\sigma_c$ is the mean standard deviation per pixel of the continuum on each side of the line, $D$ is the spectral dispersion in \AA\ per pixel, $N_{pix}$ is the number of pixels covered by the line and $EW$ is its  equivalent width.

\section{Results} \label{sec:results}
\subsection{Physical conditions of the ionized gas}

\begin{table}
     \caption{Sources of Atomic Data.}
     \label{tab:atomicData}
     \begin{tabular}{@{}l @{\hspace{0.05cm}}|p{4.2cm}|l @{\hspace{0.05cm}}|p{0.8cm}|l}
           \hline
            Ion                & Transition Probabilities       & Collision Strengths  \\
           \hline
            $\mathrm{O^+}$     & \citet{2004ADNDT..87....1F}    & \citet{2009MNRAS.397..903K}\\
            $\mathrm{O^{2+}}$  & \citet{2004ADNDT..87....1F}    & \citet{2014MNRAS.441.3028S}\\
            $\mathrm{S^+}$     & \citet{2009JPCRD..38..171P}    & \citet{2010ApJS..188...32T}\\ 
            $\mathrm{S^{2+}}$  & \citet{2009JPCRD..38..171P}    & \citet{1999ApJ...526..544T}\\ 
            $\mathrm{N^+}$     & \citet{2004ADNDT..87....1F}    & \citet{2011ApJS..195...12T}\\
            $\mathrm{Cl^{2+}}$ & \citet{1983IAUS..103..143M}    & \citet{1989AA...208..337B} \\
            \hline
 \end{tabular}
 
 \medskip 
\end{table}

\begin{table*}
   \centering
   \caption{Physical conditions of \hii\ regions in the LMC and SMC.}
   \label{tab:physical_conditions}
   \begin{tabular}{lcccccc}
        \hline
        \multicolumn{7}{c}{LMC}\\
        Parameter               & Lines       & 30 Doradus       & N44C                 & IC\,2111           &  NGC\,1714        & N11B            \\                 
        \hline                                                                                                                              
$n_\mathrm{e}\ (\mathrm{cm^{-3}})$&  \fsii    & $350 \pm 30$     & $200 \pm 150$        & $250 \pm 100$    & $350 \pm 100$   & $230 \pm 120$   \\
                                &  \foii      & $440 \pm 25$     & $200 \pm 150$        & $300 \pm 100$    & $450 \pm 100$   & $270 \pm 150$   \\
                                &  \fcliii    & $260 \pm 100$    & $440 \pm 150$        & $600 \pm 400$    & $350 \pm 200$   & $280 \pm 100$   \\
    $T_\mathrm{e}\ (\mathrm{K})$&  \foiii     & $9800 \pm 300$   & $11300 \pm 100$      & $9150 \pm 100$   & $9500 \pm 100$  & $9100 \pm 100$  \\
                                &  \fsiii     & $9000 \pm 100$   & $11400 \pm 500$      & $7950 \pm 250$   & $8000 \pm 250$  & $9800 \pm 350$  \\
                                &  \fnii      & $10000 \pm 150$  & $10500 \pm 300$      & $9750 \pm 200$   & $10200 \pm 300$ & $9700 \pm 150$  \\
                                &  \foii      & $13200 \pm 300$  & $12200 \pm 700$      & $9800 \pm 500$   & $11000 \pm 700$ & $11000 \pm 550$\\
        \hline                                                                                                                              
   \end{tabular}
   \begin{tabular}{lccccc}
        \multicolumn{6}{c}{SMC}\\
        &                                     &    N66A          &    N81               &   NGC\,456             &   N88A                          \\
        \hline                                                                                                                              
$n_\mathrm{e}\ (\mathrm{cm^{-3}})$&  \fsii    & $180 \pm 100$    & $350 \pm 150$        & $200 \pm 50$     &$2000 \pm 1000$              \\
                                &  \foii      & $190 \pm 120$    & $450 \pm 200$        & $200 \pm 20$     &$2600 \pm 1000$                     \\        
                                &  \fcliii    & $320 \pm 100$    & $400 \pm 100$        &$100^{+300}_{-100}$&$4300 \pm 400$                    \\        
    $T_\mathrm{e}\ (\mathrm{K})$&  \foiii     & $12500 \pm 100$  & $12800 \pm 150$      &$12000 \pm 100$   &    $14500 \pm 200$                 \\        
                                &  \fsiii     & $14200 \pm 550$  & $13900 \pm 550$      &    --            &    $13900 \pm 350$                \\        
                                &  \fnii      & $12000 \pm 250$  & $11900 \pm 250$      &$11200 \pm 550$   &    $13000 \pm 500$                \\        
                                &  \foii      & $13300 \pm 700$  & $12500 \pm 700$      &$16000 \pm 500$   &    $15000 \pm 1500$                \\

        \hline
   \end{tabular}
\end{table*}

The physical conditions -- electron density ($n_\mathrm{e}$) and temperature ($T_\mathrm{e}$) -- of the nebulae were derived using \textsc{pyneb} \citep{2015A&A...573A..42L} and the atomic data set shown in Table~\ref{tab:atomicData}. We determined $n_\mathrm{e}$ through the following diagnostics: \fsii\ 6731/6716, \foii\ 3726/3729, and \fcliii\ 5538/5518 and $T_\mathrm{e}$ through \foiii\ 4363/5007, \fsiii\ 6312/9069, \fnii\ 5755/6584, and \foii\ 7320+30/3726+29. We obtain direct determinations of $T_\mathrm{e}$ for the whole sample. We assumed a two-zone scheme; a high-ionization zone characterized by $T_\mathrm{e}$(\foiii) and  $n_\mathrm{e}$(\fcliii)  and a low-ionization one characterized by  $T_\mathrm{e}$(\fnii) and the weighted mean of $n_\mathrm{e}$(\fsii) and $n_\mathrm{e}$(\foii). Table~\ref{tab:physical_conditions} shows the computed values of $T_\mathrm{e}$ and $n_\mathrm{e}$ for each object.

The uncertainties in the physical conditions were computed through Monte Carlo simulations. We generated 500 random values for each diagnostic assuming a Gaussian distribution with a standard deviation equal to the associated uncertainty of the line intensities involved in the diagnostic. With those distributions, we calculated new simulated values of $T_\mathrm{e}$ and $n_\mathrm{e}$. Their associated errors correspond to a deviation of of 68 per cent -- equivalent to one standard deviation -- centred in the mode of the distribution (not in the mean). We decided to discard the standard deviation as representative of the error because if the new distributions ($T_\mathrm{e}$ and $n_\mathrm{e}$ ones) lost the symmetry property, the mean, median and mode occur at different points of the distribution and the standard deviation is not a good indicator of the error. The same procedure was used to compute the uncertainties of the ionic abundances.

\subsection{Ionic and total abundances}

The main aim of this work is the study of C and O abundances in the MCs. We calculate the ionic abundances of O$^+$ and O$^{2+}$ derived from CELs, the O$^{2+}$ and C$^{2+}$ ones obtained from RLs as well as the total abundances of O and C for all the objects of the sample. In the case of N44C, in the LMC, we also calculate the He$^+$ and He$^{2+}$ abundances because they are needed to estimate the total O/H ratio via the application of an ionization correction factor (ICF) that use those abundances (see below).

The ionic abundances from CELs were computed with \textsc{pyneb} using the atomic data set in Table~\ref{tab:atomicData}. We assumed the physical conditions of the low-ionization zone for determining the O$^+$ abundance and those of the high-ionization zone for the O$^{2+}$ one. In addition, we computed the O$^{2+}$ abundances from \oii\ RLs belonging to multiplet 1 at $\lambda \sim $ 4650 \AA. \citet{2003ApJ...595..247R} and \citet{2003MNRAS.338..687T} draw attention to the fact that under densities $n_\mathrm{e} < 10^4\ \mathrm{cm^{-3}}$ the population of the upper levels of this multiplet can be affected by departures from LTE. We used the prescriptions of \citet{2005RMxAC..23....9P} to calculate the appropriate corrections for the relative strengths between the individual \oii\ lines using the effective recombination coefficients for case B, assuming LS coupling, calculated by \citet{1994A&A...282..999S} and the $T_\mathrm{e}$ of the high-ionization zone. We calculated the C$^{2+}$ abundances from \cii\ 4267 \AA\ RL using the \cii\ effective recombination coefficients computed by \citet{2000A&AS..142...85D} for case B and  the $T_\mathrm{e}$ of the high-ionization zone. All the ionic abundances calculated in this paper are included in Table~\ref{tab:abundances}. 

For N44C, we have derived the He$^+$ abundance from the intensity of several {\hei} lines. We used {\sc pyneb} and the recent effective recombination coefficient computations by \citet{2012MNRAS.425L..28P, 2013MNRAS.433L..89P} for {\hei}, where the collisional contribution and optical depth effects in the triplet lines are included. The final adopted He$^{+}$ abundance is the weighted average of the ratios obtained from the individual lines. We detect {\heii} 4686 \AA\ line in N44C due to the high ionization degree of the nebula. We have determined the He$^{2+}$ abundance using {\sc pyneb} and the effective recombination coefficient calculated by \citet{1995MNRAS.272...41S}. The He$^+$  and He$^{2+}$ abundances of N44C are also included in Table~\ref{tab:abundances}. 

For all the objects except N44C, the O total abundances have been calculated adding the O$^{+}$ and O$^{2+}$  derived from CELs. For N44C, we compute the O total abundances adding the O$^{+}$ and O$^{2+}$ derived from CELs corrected for the O$^{3+}$ contribution adopting the ICF proposed by \citet{2014MNRAS.440..536D}. In addition, we also computed O total abundances using the O$^{2+}$ ones derived from RLs and the $\mathrm{O}^+ / \mathrm{O}^{2+}$ ratio obtained from CELs to correct for the contribution of the $\mathrm{O}^+/\mathrm{H}^+$ ratio. The total abundances of C were derived using the ICF computed by \citet{1999ApJ...513..168G} from photoionization models to correct from the unseen ionization stages of this element, mainly $\mathrm{C^+}$. The total C and O abundances are included in Table~\ref{tab:abundances}.

\begin{table*}
   \centering
   \caption{Ionic and total abundances in unit $12 + \log (\mathrm{X^{n+}/H^+)}$ of \hii\ regions in the LMC and SMC.}
   \label{tab:abundances}
   \begin{tabular}{lccccc}
        \hline
        \multicolumn{6}{c}{LMC}\\
                               &  30 Doradus     & N44C           &    IC2111      &     NGC1714    &  N11B  \\
        \hline
        \multicolumn{6}{c}{Ionic abundances}\\
        $\mathrm{O^{+}}$(CELs) &$7.70 \pm 0.04$  &$7.32 \pm 0.08$ &$8.06 \pm 0.07$ &$7.65 \pm 0.08$ &$7.98 \pm 0.06$ \\
        $\mathrm{O^{2+}}$(RLs) &$8.43 \pm 0.05$  &$8.53 \pm 0.02$ &$8.36 \pm 0.11$ &$8.47 \pm 0.08$ &$8.38 \pm 0.03$ \\
        $\mathrm{O^{2+}}$(CELs)&$8.29 \pm 0.01$  &$8.22 \pm 0.03$ &$8.18 \pm 0.04$ &$8.27 \pm 0.04$ &$8.18 \pm 0.02$ \\
        ADF(O$^{2+}$)          &$0.14 \pm 0.05$  &$0.31 \pm 0.03$ &$0.18 \pm 0.12$ &$0.20 \pm 0.09$ &$0.20 \pm 0.04$ \\
        $\mathrm{C^{2+}}$(RLs) &$7.93 \pm 0.05$  &$8.10 \pm 0.02$ &$7.99 \pm 0.01$ &$8.02 \pm 0.01$ &$7.94 \pm 0.03$ \\
        $\mathrm{C^{2+}}$(CELs)&$7.71 \pm 0.23^a$ / $7.43 \pm 0.25^b$&--&$7.98 \pm 0.25^b$   &$7.92 \pm 0.25^b$   &       --       \\
        ADF(C$^{2+}$)          &$0.22 \pm 0.23$ / $0.50 \pm 0.25$ &--&$0.01 \pm 0.25$     &$0.10 \pm 0.25$     &       --       \\
        $\mathrm{He^{+}}$      & --              &$10.8 \pm 0.01$ &      --        &     --         &    --          \\
        $\mathrm{He^{2+}}$     & --              &$10.1 \pm 0.02$ &      --        &     --         &    --          \\
       \multicolumn{6}{c}{Total abundances}\\
        C  (RLs)               &$8.03 \pm 0.05$  &$8.15 \pm 0.02$ &$8.19 \pm 0.03$ &$8.11 \pm 0.01$ &$8.11 \pm 0.03$ \\
        O  (RLs)               &$8.53 \pm 0.04$  &$8.58 \pm 0.02$ &$8.61 \pm 0.07$ &$8.56 \pm 0.07$ &$8.59 \pm 0.02$ \\
        O  (CELs)              &$8.39 \pm 0.01$  &$8.31 \pm 0.03$ &$8.43 \pm 0.04$ &$8.37 \pm 0.04$ &$8.39 \pm 0.03$ \\
  $\log (\mathrm{C/O})$ (RLs)  &$-0.50 \pm 0.06$ &$-0.43 \pm 0.03$&$-0.42 \pm 0.08$&$-0.45 \pm 0.07$&$-0.48 \pm 0.04$\\
  $\log (\mathrm{C/O})$ (CELs) &$-0.48 \pm 0.26^a$ / $-0.58 \pm 0.04^b$& --&$-0.43 \pm 0.04^b$&$-0.47 \pm 0.04^b$&     --\\

        $t^2$                  &$0.028 \pm 0.005$&$0.069 \pm 0.016$&$0.031 \pm 0.012$&$0.033 \pm 0.009$&$0.033 \pm 0.012$\\ 
        \hline
   \end{tabular}
   \begin{tabular}{lcccc}
        \multicolumn{5}{c}{SMC}\\
                               &  N66A           &    N81         &   NGC456       &   N88A              \\
        \hline
        \multicolumn{5}{c}{Ionic abundances}\\
        $\mathrm{O^{+}}$(CELs) & $7.50 \pm 0.06$ &$7.31 \pm 0.06$ &$7.57 \pm 0.13$ &$6.91 \pm 0.15$      \\
        $\mathrm{O^{2+}}$(RLs) &$8.19 \pm 0.04$  &$8.24 \pm 0.02$ &$8.18 \pm 0.13$ &$8.18 \pm 0.02$      \\
        $\mathrm{O^{2+}}$(CELs)& $7.84 \pm 0.02$ &$7.91 \pm 0.02$ &$7.89 \pm 0.02$ &$7.89 \pm 0.03$      \\
        ADF(O$^{2+}$)           &$0.35 \pm 0.13$ &$0.33 \pm 0.11$ &$0.29 \pm 0.13$ &$0.29 \pm 0.10$      \\
        $\mathrm{C^{2+}}$(RLs) &$7.54 \pm 0.04$  &$7.58 \pm 0.03$ &$7.60 \pm 0.17$ &$7.63 \pm 0.02$      \\
        $\mathrm{C^{2+}}$(CELs)&$7.09 \pm 0.15^b$&$7.15 \pm 0.15^b$&    --         &$7.25 \pm 0.04^a$        \\
        ADF(C$^{2+}$)          &$0.45 \pm 0.15$  &$0.43 \pm  0.15$ &    --          &$0.38 \pm 0.04$      \\
        
        \multicolumn{5}{c}{Total abundances}\\
        C (RLs)                &$7.68 \pm 0.05$  &$7.67 \pm 0.04$ &$7.74 \pm 0.17$ &$7.68 \pm 0.02$      \\
        O (RLs)                &$8.35 \pm 0.03$  &$8.34 \pm 0.02$ &$8.36 \pm 0.09$ &$8.22 \pm 0.02$      \\
        O (CELs)               &$8.00 \pm 0.02$  &$8.01 \pm 0.02$ &$8.06 \pm 0.05$ &$7.94 \pm 0.03$      \\

  $\log (\mathrm{C/O})$ (RLs)  &$-0.67 \pm 0.06$ &$-0.67 \pm 0.04$&$-0.62 \pm 0.19$&$-0.54 \pm 0.03$     \\
  $\log (\mathrm{C/O})$ (CELs) &$-0.86 \pm 0.02^b$ &$-0.91 \pm 0.02^b$&    --          &$-0.72 \pm 0.17^a$     \\

  $t^2$                  &$0.090 \pm 0.019$&$0.091 \pm 0.017$&$0.075 \pm 0.014$&$0.107 \pm 0.027$  \\       
         
        \hline
   \multicolumn{5}{l}{$^a$ $\mathrm{C^{2+}}$ and C/O abundances derived from CELs in the UV by \citet{1995ApJ...443...64G}. } \\
   \multicolumn{5}{l}{$^b$ $\mathrm{C^{2+}}$ and C/O abundances derived from CELs in the UV by \citet{1982ApJ...252..461D}.}\\
   \multicolumn{5}{l}{We assume the error of $\mathrm{C/H}$ ratios given in \citet{1982ApJ...252..461D} as representative  }\\
   \multicolumn{5}{l}{of the error of $\mathrm{C^{2+}/H^+}$ ratios since this is not given in the paper.}\\
   \multicolumn{5}{l}{The true errors in C/O ratios from UV CELs are probably larger than}\\
   \multicolumn{5}{l}{the quoted  values according to the authors.}

   \end{tabular}

\end{table*}

\section{Discussion}\label{sec:discuss}

\subsection{Nebular abundances and the abundance discrepancy}

\begin{figure}
 \centering
  \includegraphics[]{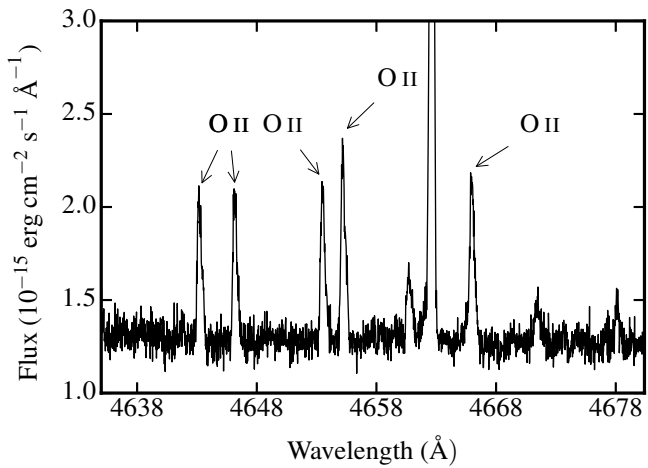}
 \caption{Section of the UVES spectrum of the \hii\ region N11B of the SMC around the RLs of multiplet 1 of \oii. Absorption lines due to scattered starlight are not detected in the spectrum.}
 \label{fig:spec_n11}
\end{figure}

Table~\ref{tab:abundances} shows the values of the ADF(O$^{2+}$) found in our sample. The measured ADFs are systematically larger in the SMC than in the LMC. The weighted mean of the ADF(O$^{2+}$) is $0.25 \pm 0.02\ \mathrm{dex}$ for the LMC and $0.32 \pm 0.06\ \mathrm{dex}$ for the SMC. The small dispersion of the ADF values within each galaxy is remarkable. Several works in the literature \citep[see e.~g.][and references therein]{2007ApJ...670..457G, 2009ApJ...700..654E} have reported that values of the ADF in \hii\ regions are typically between 0.1 and 0.35 dex. The ADFs obtained in the SMC are at the higher end of these typical values. \citet{2014MNRAS.443..624E} and \citet[][]{2016MNRAS.458.1866T} have  found ADFs of about 0.3-0.4 dex in the case of star-forming dwarf galaxies and \hii\ regions in the spiral galaxy NGC\, 300, respectively. \citet{2003MNRAS.338..687T} determined the ADF(O$^{2+}$) for 30 Doradus and N11B in the LMC and for N66 in the SMC. Since for N66 their value is consistent with our determination, the ADF(O$^{2+}$) obtained by these authors for 30 Doradus and N11B are larger than ours. In particular, \citet{2003MNRAS.338..687T} derive an ADF(O$^{2+}$) of about 0.91 dex for N11B. The reason of this unusually large value is due mainly because they measure a very large intensity of the 3d$-$4f \oii\ 4089 \AA\ line that also gives a very large O$^{2+}$ abundance. \citet{2003MNRAS.338..687T} average the O$^{2+}$/H$^+$ ratio obtained from this line and that obtained from those of multiplet 1 of \oii\ to derive the final O$^{2+}$ abundance adopted for N11B. However, we only find a faint noisy feature at the expected wavelength of the line in our spectrum. Even this feature seems to be affected by a ghost due to charge transfer from the bright \foiii\ 4959 \AA\ line that lies in a nearby spectral order. With the intensity we measure for this feature --that may be attributed to \oii\ 4089 \AA--, we derive an abundanc of the order of that obtained from the lines of multiplet 1 of \oii. Other reason of the larger values of ADF(O$^{2+}$) obtained by \citet{2003MNRAS.338..687T} for N11B is that they assumed a substantial contamination by scattered starlight in the nebular spectrum and apply important upward corrections to the intensity of the nebular \oii\ RLs. In this sense, as it can be seen in Fig.~\ref{fig:spec_n11}, our much higher resolution spectrum does not show any trace of the presence of such absorption features, so the corrections do not seem to be necessary. \citet{2003MNRAS.338..687T} obtain an ADF(O$^{2+}$) = 0.33 dex when they only consider the \oii\ RLs of multiplet 1 and no contamination by scattered starlight. This value is much more consistent with our determination of 0.20 dex for N11B. Therefore, we have enough evidence to dismiss the presence of large ADF(O$^{2+}$) values in this object.

We also calculated the ADF(C$^{2+}$) using the C$^{2+}$ abundances derived from CELs in the UV by \citet{1995ApJ...443...64G} and \citet{1982ApJ...252..461D}. The results are shown in Table~\ref{tab:abundances}. These ADFs show a larger dispersion than for O$^{2+}$ in both galaxies. There are several reasons that could be affecting the precision of the ADF(C$^{2+}$) values: i) C$^{2+}$ abundances derived from CELs in the UV are strongly affected by uncertainties in the reddening correction \citep{1995ApJ...443...64G, 1999ApJ...513..168G}; and ii) HST FOS and IUE apertures are very different from the long-slit or echelle spectra therefore the ADFs(C$^{2+}$) are based on different nebular volumes covered by the UV and optical observations, while the ADFs(O$^{2+}$) refer to the same volume of gas. The weighted mean values of the ADF(C$^{2+}$) are $0.15 \pm 0.24$ dex for the LMC and $0.42 \pm 0.08$ dex for the SMC. Again, the objects of the SMC show larger values than those of the LMC.

\begin{figure*}
 \centering
  \includegraphics[width=0.49\textwidth]{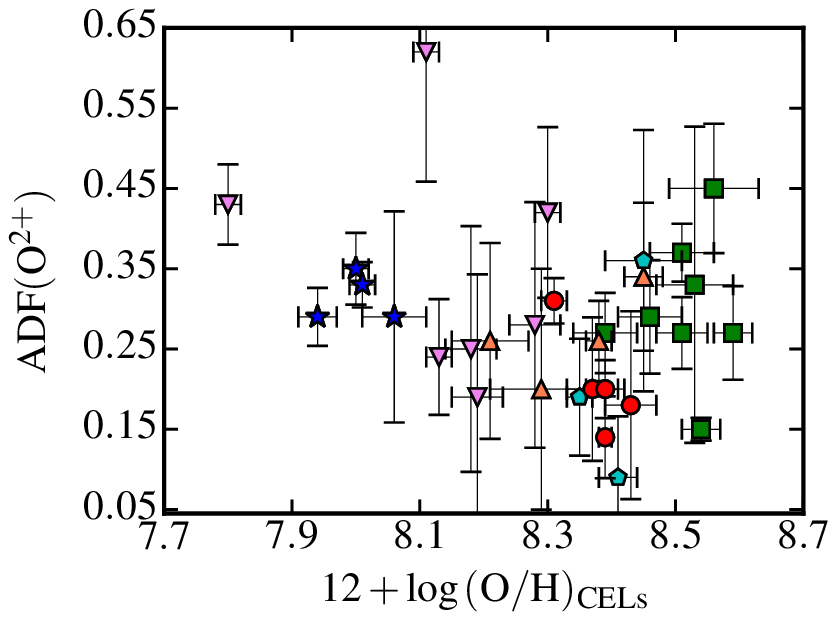}
  \includegraphics[width=0.49\textwidth]{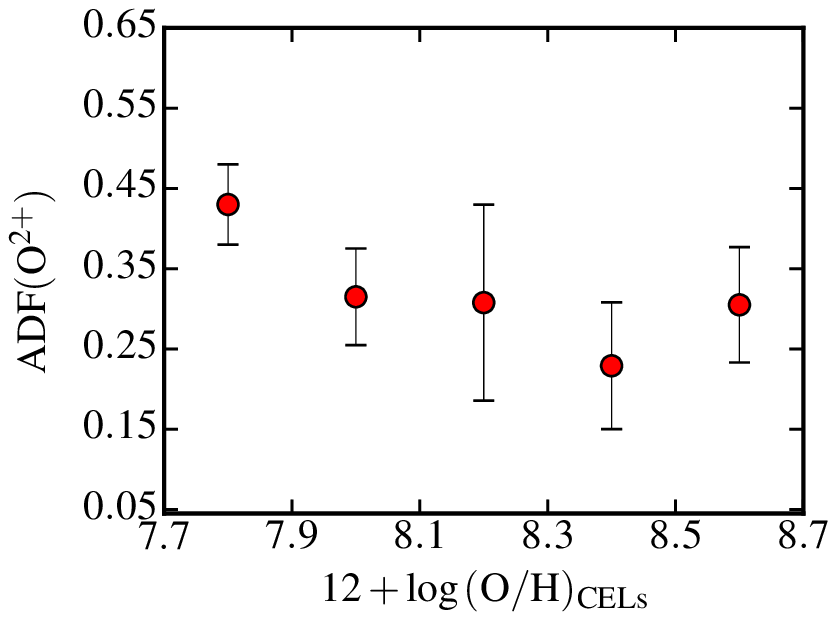}
  \includegraphics[width=0.49\textwidth]{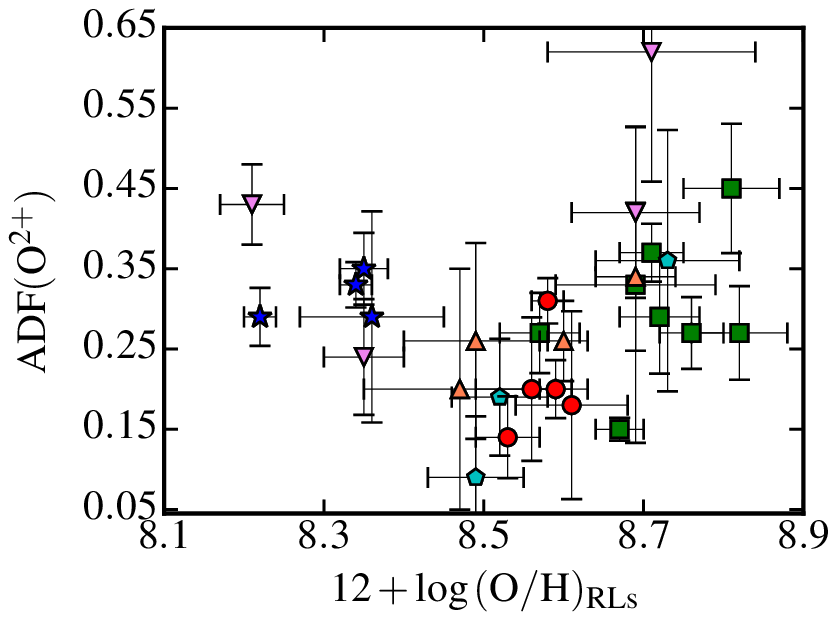}
  \includegraphics[width=0.49\textwidth]{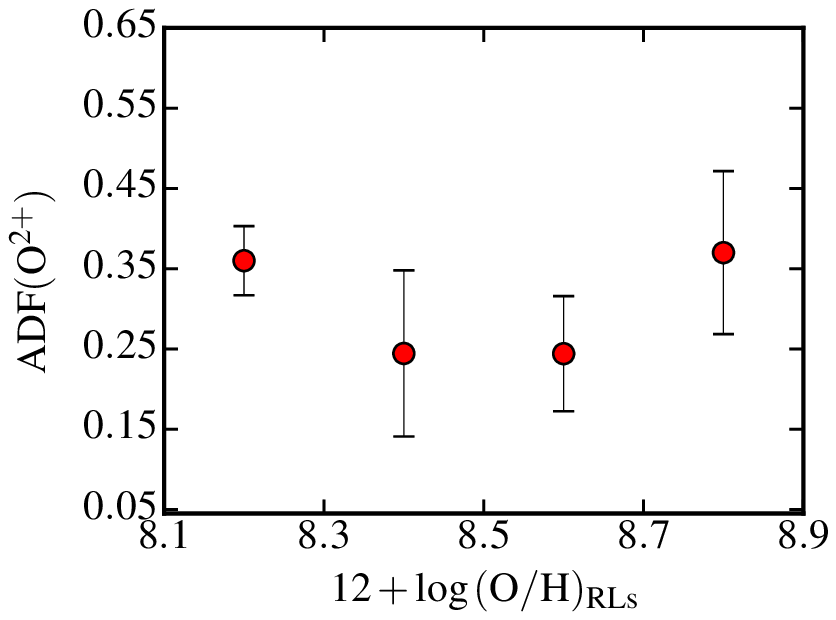}
 \caption{ADF(O$^{2+}$) {\it versus} O/H ratio. In the left-hand panels we show data for \hii\ regions in different galaxies -- SMC (blue stars; this work), LMC (red circles; this work), M33 (orange triangles; \citet{2016MNRAS.458.1866T}),  MW \citep[green squares; ][]{2005MNRAS.362..301G, 2006MNRAS.368..253G, 2007RMxAA..43....3G, 2004ApJS..153..501G, 2004MNRAS.355..229E, 2013MNRAS.433..382E}, M101 \citep[cyan pentagons][]{2009ApJ...700..654E} and dwarf galaxies \citep[pink down-facing triangles][]{2007ApJ...656..168L, 2014MNRAS.443..624E}. The O abundances are derived either from CELs (upper panels) or from RLs (lower panels). Right-hand panels represent the average of ADF(O$^{2+}$) for different abundance bins.}
 \label{fig:adf_oxygen}
\end{figure*}

In Table~\ref{tab:abundances} we also compare the C/O ratios derived from RLs and from CELs in the UV. The C/O ratios obtained from RLs are almost identical to those from CELs in the case of the LMC but not in the SMC, where log(C/O) from CELs are $~0.2$ dex systematically lower than those from RLs\footnote{This behaviour may be explained in the framework of the temperature fluctuations paradigm. UV CELs of C$^{2+}$ are more affected by such fluctuations than optical CELs of O$^{2+}$, and ionic abundances determined from CELs should produce lower abundances with respect to the ``true" ones in the case of C$^{2+}$ \citep[see][]{2007ApJ...670..457G}.}. Only for the region N88A, the C/O ratios determined from both kinds of lines are in agreement within the errors. \citet{1982ApJ...252..461D} state in their paper that the true errors of the mean C/O ratios for the rest of the objects of the SMC are probably several times larger than the nominal values they give in their work. Therefore, this apparent disagreement may not be so large. Several previous works indicate that the C/O ratios determined from CELs and RLs are fairly similar in \hii\ regions \citep{2014MNRAS.443..624E} as well as in planetary nebulae \citep[e.g.][]{1998ApJ...495..328M, 2000MNRAS.312..585L}. 

In addition, in Table~\ref{tab:abundances} we included the temperature fluctuation parameter ($t^2$) derived following the formalism developed by \citet{1969BOTT....5....3P}. As in the case of the ADF(O$^{2+}$),  \hii\ regions in LMC show a remarkable small dispersion of their $t^2$, excluding N44C. For the \hii\ regions in the SMC, $t^2$ values are systematically much higher but the dispersion is also small. 

\citet{2007ApJ...670..457G} explored the possible dependence of the ADF with different properties of \hii\ regions. Those authors reported that the ADF seems to be independent of metallicity;\footnote{In nebular astrophysics, the O abundance is used as proxy of the metallicity, as this is the most abundant element after H and He.} however, their sample comprises only one object with metallicity lower than $12 + \log(\mathrm{O/H}) = 8.0$. The new information we are providing with accurate ADFs for low-metallicity \hii\ regions of the SMC -- in addition to the results for some objects studied by \citet{2007ApJ...656..168L} and \citet{2014MNRAS.443..624E} -- enables us to extend the analysis performed by  \citet{2007ApJ...670..457G} to lower metallicities. Fig.~\ref{fig:adf_oxygen} shows the ADF(O$^{2+}$) determined in this work and other previous papers as a function of the total O abundances derived from CELs (upper panels) and from RLs (lower panels). Apart from the data for the \hii\ regions of the MCs, we include results for M33 \citep{2016MNRAS.458.1866T},  MW \citep{2005MNRAS.362..301G, 2006MNRAS.368..253G, 2007RMxAA..43....3G, 2004ApJS..153..501G, 2004MNRAS.355..229E, 2013MNRAS.433..382E}, M101 \citep{2009ApJ...700..654E} and star-forming dwarf galaxies \citep{2007ApJ...656..168L, 2014MNRAS.443..624E}. In the right-panels, we computed the mean of ADF(O$^{2+}$) values for different metallicity bins of the O/H ratio determined either from CELs or RLs. We defined bins 0.2\,dex wide. The first bin begins at $12 + \log(\mathrm{O/H}) = 7.7$ from CELs and $12 + \log(\mathrm{O/H}) = 8.1$ from RLs. We computed the mean of the ADF(O$^{2+}$) for all the objects that fall into each bin and we plotted these values in the middle point of each bin.

Although the ADF(O$^{2+}$) values estimated for an important fraction of the objects represented in Fig.~\ref{fig:adf_oxygen} have large uncertainties, the distribution of the points  suggests a complex behaviour. We can see that the ADF(O$^{2+}$) values show an apparent minimum at 12 + log(O/H) $\sim$ 8.4 or 8.5 -- when determined from CELs or RLs, respectively -- and that the mean value of this quantity increases when the O/H becomes higher or lower. This complex shape of the of ADF(O$^{2+}$) distribution is clearer when O/H ratios determined from RLs are considered as baseline for O abundances. If real, the explanation of this behaviour is a difficult task. From our point of view, the accurate values of the ADF(O$^{2+}$) we have obtained for the {\hii} regions  of the MCs has been the main factor that has permitted to find this possible metallicity-dependent behaviour of the ADF.  The high S/N ratio of the spectra presented in this paper, the high ionization degree of the {\hii} regions and the absence of important underlying stellar contribution have contributed to the high quality and low uncertainty of the ADF(O$^{2+}$) determinations in the {\hii} regions of the MCs. In the case of Galactic {\hii} regions we obtain a high dispersion of ADF(O$^{2+}$) determinations and a tendency to larger values for the different individual objects. Considering the closeness of the Galactic nebulae, the dispersion may be due to the fact that we obtain the spectra from bright very small areas of the objects and that the ADF(O$^{2+}$) calculated for the nebulae may be affected by the local conditions of the gas, which could not be representative of the whole nebula. As it has been proven by \citet{ 2008ApJ...675..389M, 2009MNRAS.395..855M, 2011MNRAS.417..420M} the presence of localized high-velocity flows and/or high density gas -- with higher surface brightness -- in the line of sight of the observed aperture can produce higher values of the ADF(O$^{2+}$). In any case, although the local effects may be affecting the position of some Galactic objects in Fig.~\ref{fig:adf_oxygen} we cannot exclude the presence of a trend to higher ADF(O$^{2+}$) when 12 + log(O/H) is larger than about 8.4-8.5. Interestingly, the photoionization models presented by \citet[][their fig. 32]{2013ApJS..208...10D}, calculated for a $\kappa$ parameter -- that describes the extent to which the energy distribution of the free electron departs from thermal  equilibrium distribution -- of about 20 predict an increment of the ADF(O$^{2+}$) for 12 + log(O/H) larger than 8.50 but constant values for lower O/H ratios. In Fig.~\ref{fig:adf_oxygen}, the trend for the low-metallicity objects seems to be clearer than the high-metallicity ones.  Because we include more distant objects in this zone of the diagram, their spectra correspond to apertures that encompass a large fraction of the nebula and, therefore, are more representative of the global emission and the aforementioned localized phenomena that can affect the ADF may be washed out. However, another mechanism can be acting in contributing to the large ADF(O$^{2+}$) values reported in some giant {\hii} regions belonging to star-forming dwarf galaxies (pink down-facing triangles shown in Fig.~\ref{fig:adf_oxygen}). \citet{2016MNRAS.460.4038E} speculate that large ADFs in {\hii} regions might be produced by the presence of shocks due to large-scale interaction of stellar winds with the ionized gas that may dominate the kinematics of these complex and huge nebulae. This suggestion is based on the results obtained by \citet{2016MNRAS.460.4038E} and \citet{2014ApJ...785..100M}, who determined the ADF(O$^{2+}$) in several slit positions of NGC~6888 and NGC~7635, two Galactic ring nebulae associated with evolved massive stars. The mean ADF(O$^{2+}$) of NGC~6888 and NGC~7635 is 0.50$\pm$0.06 dex and 0.38$\pm$0.12, respectively, being both values larger than the mean values of normal {\hii} regions with the same O abundance but similar to the high ADF(O$^{2+}$) values found for star-forming 
dwarf galaxies.

\begin{figure*}
 \centering
  \includegraphics[width=0.49\textwidth]{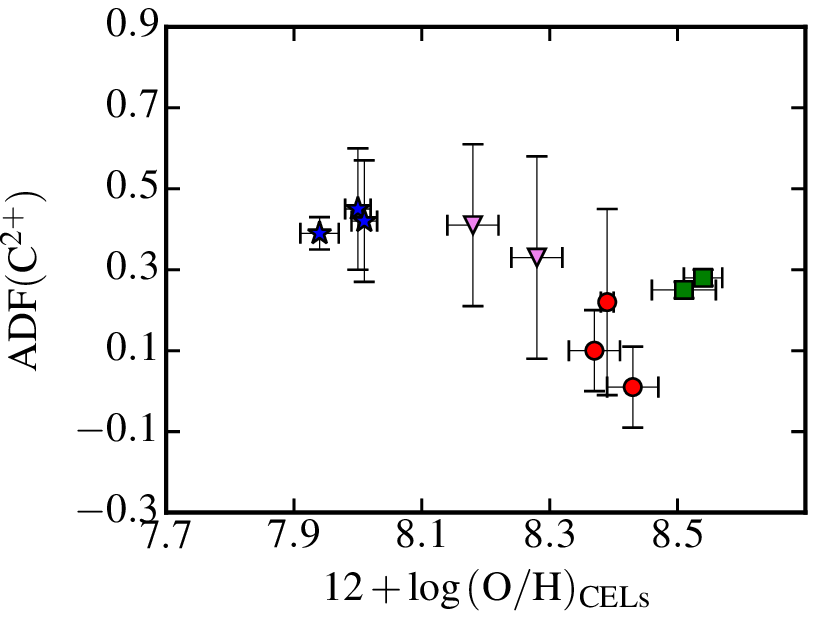}
  \includegraphics[width=0.49\textwidth]{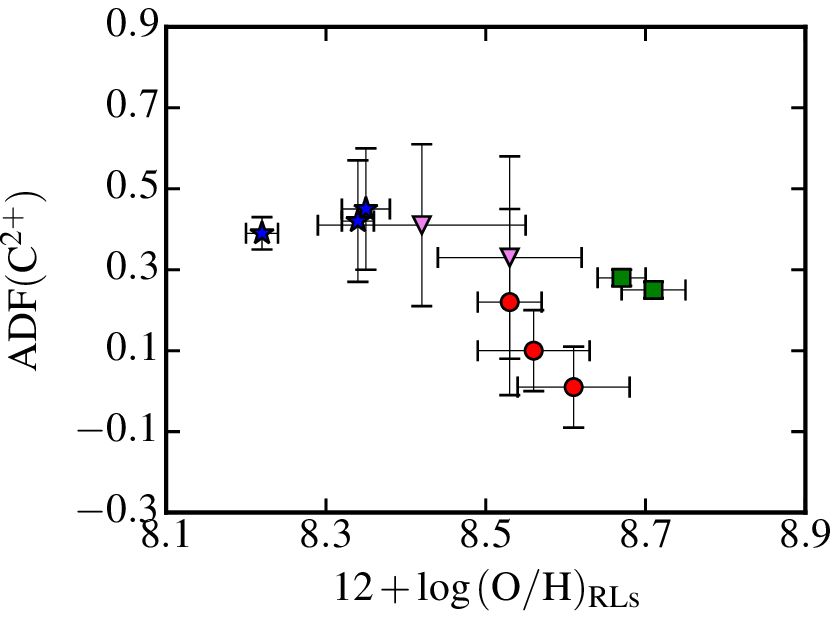}
 \caption{ADF(C$^{2+}$) {\it versus} O/H ratio for \hii\ regions in different galaxies: SMC (blue stars; this work), LMC (red circles; this work), MW \citep[green squares; ][]{2007ApJ...670..457G}, and NGC\,5253 \citep[pink down-facing triangles; ][]{2007ApJ...656..168L}. Left panel: O/H ratio determined from CELs, right panel: O/H ratio from RLs.}
 \label{fig:adf_carbon}
\end{figure*}

Fig.~\ref{fig:adf_carbon} contains the ADF(C$^{2+}$) as a function of the total O abundances determined from CELs and RLs (left and right panel, respectively) for the \hii\ regions where we have determinations of C$^{2+}$ abundances from CELs available from the literature. As it can be seen, Fig.~\ref{fig:adf_carbon} shows an apparent trend to slightly larger ADF(C$^{2+}$) at lower abundances. Unfortunately, there are only two determinations of the ADF(C$^{2+}$) for metal-rich Galactic {\hii} regions. However, their values are larger than those of the objects with 12 + log(O/H) $\sim$ 8.4 or 8.5, in agreement with the general behaviour seen in Fig.~\ref{fig:adf_oxygen}. 
In any case, there is much reservation about this point because the computation of the ADF(C$^{2+}$) involves more uncertainties than that of ADF(O$^{2+}$). The line intensity ratios of the CELs and RLs from which we derive the abundances used to estimate ADF(C$^{2+}$) have been taken with different instruments and with different slit sizes, therefore they can be very much affected by aperture effects between the UV and optical observations. This is especially dramatic in the observations of \hii\ regions in the Milky Way -- which correspond to the Orion Nebula and M8 -- owing to the large angular size of the objects and the small area covered by spectroscopical observations.

\subsection{Comparison between nebular and stellar O abundances}

An interesting test to explore the abundance discrepancy problem is to compare the abundances derived from CELs and RLs in \hii\ regions with those determined in early B-type stars located in their vicinity. The O abundance in the photospheres of early B-type stars -- which is not expected to be affected by stellar evolution effects --  should reflect the present-day chemical composition of the interstellar material in the regions where they were formed and in which they are located.

\begin{figure*}
 \centering
  \includegraphics[width=0.49\textwidth]{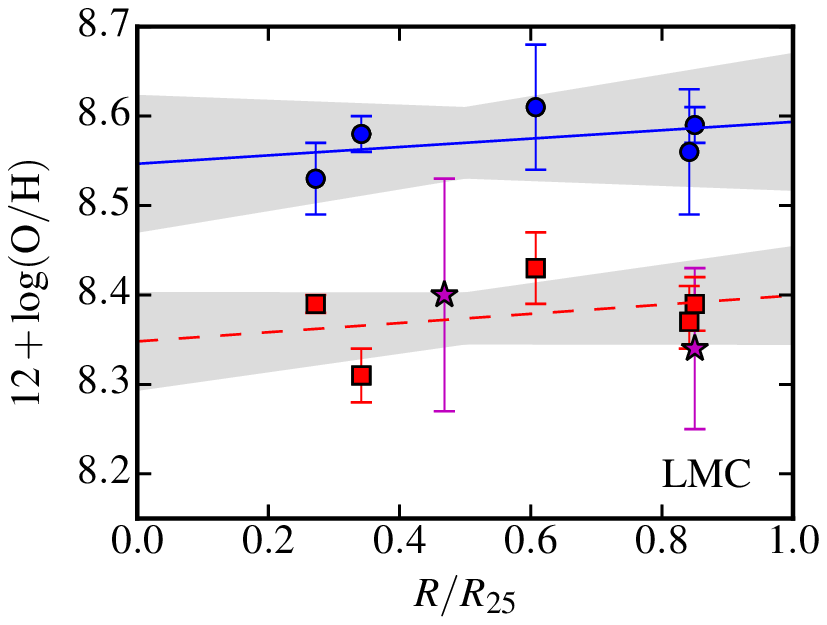}
  \includegraphics[width=0.49\textwidth]{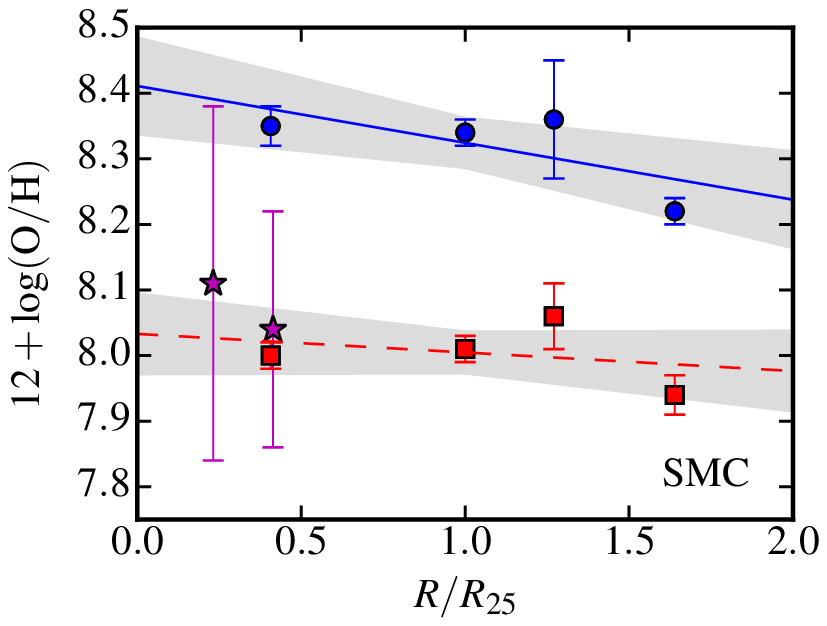}
 \caption{Spatial distribution of the O abundances as a function of the fractional galactocentric distance ($R$/$R_{25}$) for the LMC (left-hand panel) and the SMC (right-hand panel) determined from CELs (red squares and dashed line) and RLs (blue circles and continuous line) compared with the results obtained by \citet{2009A&A...496..841H} for B-type stars (magenta stars). The error bars of the stars correspond with the standard deviation of the stellar abundances included for each star-forming regions (53 and 78 stars in each cluster of the LMC and 31 and 40 stars in of the SMC ones) determined by \citet{2009A&A...496..841H}. The grey areas trace all the radial gradients compatible with the uncertainties of our least-squares linear fits computed through Monte Carlo simulations (see Section \ref{sec:radial_abun}).}
 \label{fig:stars}
\end{figure*}

\begin{figure}
 \centering
  \includegraphics[]{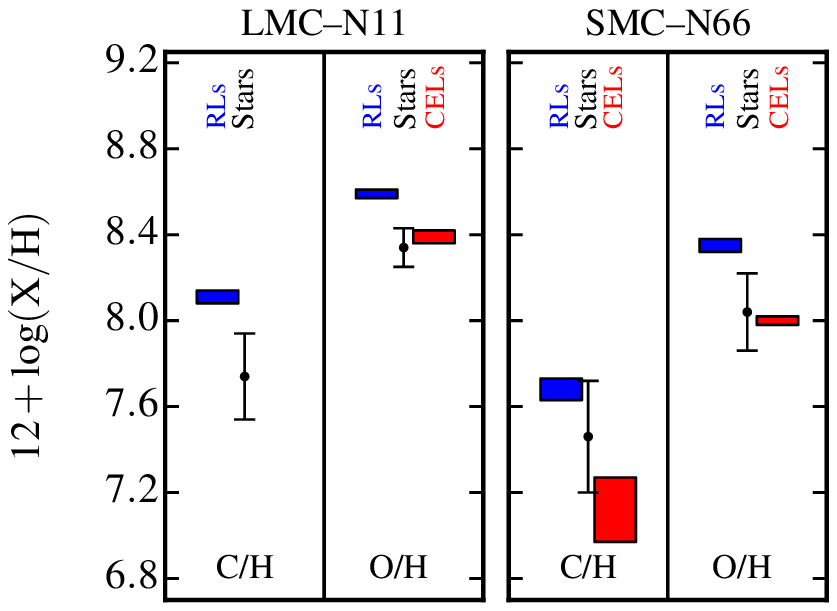}
 \caption{C and O abundances of two \hii\ regions of this work obtained from CELs (red rectangles) and RLs (blue rectangles) and B-type stars belonging to the associated clusters \citet{2009A&A...496..841H} (black dots). The height of each rectangle corresponds to the abundance uncertainty. The vertical black line represents standard deviation of the stellar abundances.}
 \label{fig:adp}
\end{figure}

In Fig.~\ref{fig:stars}, we have compared our derived O abundances for \hii\ regions with those determined by \citet{2009A&A...496..841H} for B-type stars in young clusters of the LMC and SMC. The results are plotted  for the LMC (left-hand panel) and SMC (right-hand panel). Moreover, because some of the clusters studied by \citet{2009A&A...496..841H} are associated with {\hii} regions of our sample: N11 in the LMC and N66 in the SMC, we compare the different abundance data for these objects in Fig.~\ref{fig:adp}. In this figure, we include the C and O nebular abundances derived from CELs  \citep[red rectangles, C abundances derived by ][]{1982ApJ...252..461D} and RLs (blue rectangles) and the stellar abundances \citep[determined by ][ black dots and error bars]{2009A&A...496..841H} for N11 in the LMC (left-hand panels) and N66 in the SMC (right-hand panels). The height of the nebular rectangles corresponds to the uncertainties of the abundances given in Table~\ref{tab:abundances}. The error bars for the stellar data  correspond to the standard deviation of the stellar abundances included for each star-forming region (53 stars for N11 and 31 for N66). From the results shown in Fig.~\ref{fig:stars} and Fig.~\ref{fig:adp}, we can conclude that O abundances derived for B-type stars agree better with nebular ones derived from CELs. Moreover, the nebular abundances should be increased by 0.08 dex in the most metal-poor \hii\ regions \citep{2010ApJ...724..791P} and about 0.12 dex in metal-rich ones \citep{2009MNRAS.395..855M} owing to the O embedded in dust grains. In this way, the gas+dust O abundances derived from CELs are still consistent with the stellar ones but the values obtained from RLs become even more discrepant. In the case of C, for N11 in the LMC, the nebular abundance derived from RLs is much higher than the mean stellar one. Unfortunately, C abundances from CELs are not available for this \hii\ region. For N66 in the SMC, C abundances derived from RLs and CELs are both in agreement with the stellar ones owing to the large uncertainties of stellar determinations. The amount of C embedded in dust grains is unknown, but if we assume the value 0.10 dex computed for the Orion nebula \citep{1998MNRAS.295..401E} the C abundance determined from RLs would be too high and in disagreement with the stellar one. 

Previous comparisons between stellar and nebular abundances in other galaxies show contradictory results. The studies in the Milky Way carried out by  \citet{2012A&A...539A.143N, 2011BSRSL..80..255S} and \citet{2014A&A...571A..93G} found that stellar O abundances show better agreement with nebular O/H ratios based on RLs. In the case of M31, \citet{2012MNRAS.427.1463Z} found that nebular O abundance determinations based on CELs are 0.3 dex lower than those derived by \citet{2002A&A...395..519T} and \citet{2001MNRAS.325..257S} for B supergiants and for AF stars by \citet{2000AAS...196.4011V}. Very recently \citet{2016ApJ...830...64B} compared the metallicities of A-type supergiants and \hii\ regions in the metal-rich spiral galaxy M83, finding that in general the CELs (as indicated by strong-line methods directly calibrated from auroral line detections) tend to underpredict the stellar metallicities at the high end of the abundance scale. Nevertheless, they also showed that for some individual, metal-rich (around 1.9 $\times$ solar) \hii\ regions the CELs  can provide metallicities that are in very good agreement with those from the supergiant stars. \citet{2016MNRAS.458.1866T} compared O abundances derived from CELs and RLs in \hii\ regions of the low-mass spiral galaxies M33 and NGC\,300 with the O abundances derived in B supergiants by \citet{2005ApJ...635..311U} and \citet{2005ApJ...622..862U}, respectively. These authors found that the stellar O/H ratios are in better agreement with the nebular abundances calculated using RLs in the case of M33, while in the case of NGC\,300 the CELs provide better consistency with the supergiants. This latter finding confirms the result obtained by \citet{2009ApJ...700..309B} from a comparison between nebular and stellar metallicities.

\begin{figure}
 \centering
  \includegraphics[]{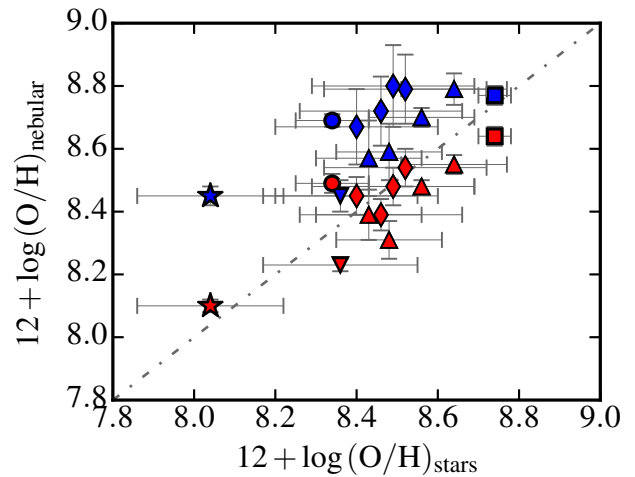}
 \caption{Comparison of the O abundance determined from young -- B or A type -- supergiant stars with those obtained from {\hii} regions located in the same star-forming region or at the same galactocentric distance in the same galaxy. We include data of the Orion Nebula and its associated cluster (squares), the star-forming region N11 in the LMC (circles), N66 in the SMC (stars), the {\hii} region Hubble V in NGC\,6822 (down-facing triangles), and several {\hii} regions in M33 (triangles) and NGC\,300 (diamonds). Nebular data have been increased 0.1 dex to correct for dust depletion. Red symbols represent nebular O/H ratios determined from CELs and blue ones O/H ratios determined from RLs. The dot-dashed line represents the 1:1 relation.}
 \label{fig:stars_vs_neb}
\end{figure}

\citet{2016ApJ...830...64B} suggested that nebular abundances determined either from CELs or RLs can display different levels of agreement with the supergiant work, depending on the metallicity regime. Their fig. 11 displays the difference between stellar and nebular abundances -- the latter corrected for dust depletion -- as a function of metallicity for 15 galaxies. For six of these galaxies these authors included nebular abundances derived from both CELs and RLs. They found that the RL-based nebular metallicities agree with the stellar metallicities better than the CELs in the high-abundance regime, but that this situation reverses at low abundance values. This result appears to be robust, despite some uncertainties. For example, as \citet{2016ApJ...830...64B} pointed out, only in the case of the Orion nebula, included in their plot, stellar and nebular abundances refer to the same star-forming region. We do not expect the adoption of average galactic abundances to affect the comparison made by these authors in the case of dwarf galaxies, thanks to the virtually homogeneous chemical composition of these systems. For the spirals, in which instead the abundances decrease with galactocentric radius, \citet{2016ApJ...830...64B} used the central metallicities as representative for each galaxy when calculating the difference between stellar and nebular values. In these systems (except M83) they thus used the known radial abundance gradients to obtain the central metallicities. In addition, an additional concern arises from the fact that spectral features belonging to a variety of heavy elements are used in the case of the A-type supergiants in order to estimate the stellar metallicities, rather than just oxygen lines. In Fig.~\ref{fig:stars_vs_neb} we have made a similar exercise than in fig. 11 of \citet{2016ApJ...830...64B} but including data for individual {\hii} regions with high quality nebular CELs and RLs determinations and compare with O/H ratios determined from young supergiant stars located in the same star-forming regions or galactocentric distance in the same galaxy. The squares correspond to data for the Orion Nebula and B-type stars of its associated cluster \citep{2004MNRAS.355..229E, 2011A&A...526A..48S}. For the MCs we include data for the ionized gas (this work) and B-type stars \citep{2009A&A...496..841H} of the star-forming regions N11 (LMC, circles) and N66 (SMC, stars). In the case of the data points corresponding to M33 (triangles) and NGC\,300 (diamonds) we have taken the nebular O abundances of the {\hii} regions IC~132, NGC~588, R20, R23, R14, R2 \citep{2016MNRAS.458.1866T}, NGC~595 and NGC~604 \citep{2009ApJ...700..654E},  and the stellar ones have been estimated from the radial O abundance gradient determined by \citet{2005ApJ...635..311U} and \citet{2005ApJ...622..862U} from the spectra of B-type supergiants, evaluated at the galactocentric distances of each {\hii} region. The down-facing triangles represent values for the dwarf irregular galaxy NGC~6822, the nebular data are taken from \citet{2014MNRAS.443..624E} and the stellar ones from the spectral analysis of two A-type supergiant stars performed by \citet{2001ApJ...547..765V}. From the figure, one can conclude that while in the case of M33 and NGC~6822 both kinds of lines give nebular abundances consistent with the stellar ones, in the other cases we obtain contradictory results. The nebular O abundances determined from RLs in the Orion Nebula are the ones consistent with stellar determinations while in the MCs and NGC\,300 the determinations based on CELs are the only ones consistent with the stellar abundances. This seems to be qualitatively consistent with the result obtained by \citet{2016ApJ...830...64B}. In the case of NGC~6822, \citet{2016ApJ...830...64B} used the metallicities -- Fe/H ratios -- estimated by \citet{2015ApJ...803...14P} for a sample of red supergiants for deriving the O/H ratio of this galaxy, obtaining a 12+log(O/H) = 8.08 $\pm$ 0.21 for this object instead of the value of 8.36 $\pm$ 0.19 obtained by \citet{
2001ApJ...547..765V}. Using this value the behaviour of NGC~6822 in the diagram would  become more similar to that of the other low-metallicity objects in the MCs -- CELs determinations more consistent with stellar abundances. However, for deriving the O/H ratio from the metallicities of the red supergiants, \citet{2016ApJ...830...64B} assume a solar O/Fe ratio and there are evidences that such ratio may be sub-solar at low metallicities \citep[e.g][]{2014ApJ...788...64G} in the case of IC\,1613.

\subsection{The spatial distribution of O and C abundances in the MCs} \label{sec:radial_abun}

\begin{table}
     \caption{Radial abundance gradients for the LMC and SMC.}
     \label{tab:fits}
     \begin{tabular}{@{}lccc}
           \hline
                                        & Lines  &intercept (n)             &slope (m)   \\
                                        &        &$\Big[ \mathrm{dex} \Big]$&$\Big[ \mathrm{dex}\  (R/R_{25})^{-1} \Big]$\\
           \hline
           \multicolumn{4}{c}{LMC}                                                                               \\
          
           $12 +  \log(\mathrm{O/H})$   & CELs   &  $8.35 \pm 0.03$ &    $0.05 \pm 0.05$\\
                                        & RLs    &  $8.55 \pm 0.04$ &    $0.04 \pm 0.07$\\
           $12 + \log(\mathrm{C/H})$    & RLs    &  $8.08 \pm 0.05$ &    $0.06 \pm 0.07$\\ 
           $ \log(\mathrm{C/O})$        & RLs    &  $-0.46\pm 0.06$ &    $0.01 \pm 0.10$\\ 
           \multicolumn{4}{c}{SMC}                                                      \\
           $12 +  \log(\mathrm{O/H})$   & CELs   &  $8.03 \pm 0.03$ &    $-0.03 \pm 0.03$\\
                                        & RLs    &  $8.41 \pm 0.04$ &    $-0.09 \pm 0.04$\\
           $12 + \log(\mathrm{C/H})$    & RLs    &  $7.68 \pm 0.06$ &    $0.01 \pm 0.06$\\ 
           $\log(\mathrm{C/O})$         & RLs    &  $-0.73\pm 0.07$ &    $0.09 \pm 0.07$\\ 
            \hline
 \end{tabular}
 
 \medskip 
\end{table}

We have studied the spatial distribution of O and C abundances as a function of the fractional galactocentric distance ($R$/$R_{25}$) in the LMC and SMC in order to explore the presence of radial abundance gradients in both galaxies. We performed least-squares linear fits to the abundance determinations obtained using both kinds of lines, CELs and RLs. For each \hii\ region, $R/R_{25}$ has been taken from Table~\ref{tab:information} and the O and C abundances from Table~\ref{tab:abundances}. We compute the radial gradients up to 5 kpc in both galaxies. We give the intercept and the slope -- in $\mathrm{dex}\ (R/R_{\mathrm 25})^{-1}$ -- of each fit in Table~\ref{tab:fits}. The uncertainties of the gradients were computed through 100 Monte Carlo simulations. The O fits are plotted in Fig.~\ref{fig:stars} (LMC: left-hand panel; SMC: righ-hand panel). Within the uncertainties, we find that the slope of the O/H gradient is almost the same whether we use CELs or RLs. In previous works, \citet{2005ApJ...618L..95E, 2009ApJ...700..654E} and \citet{2016MNRAS.458.1866T} also reported similar slopes of the O/H gradients - independently of the kinds of lines used to derive the abundances - for the Milky Way, M101, NGC\,300 and M33. 

\begin{figure*}
 \centering
  \includegraphics[width=0.48\textwidth]{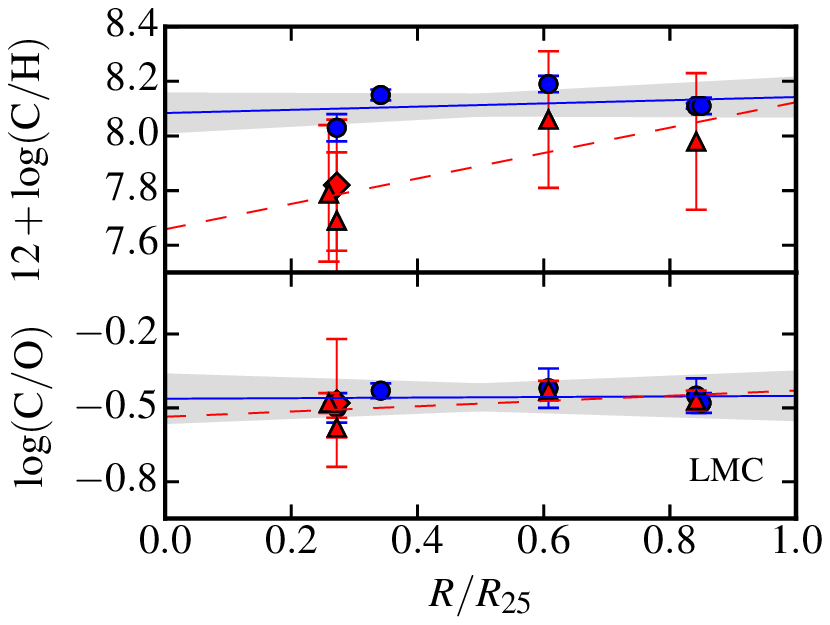}
  \hspace{2mm}
  \includegraphics[width=0.48\textwidth]{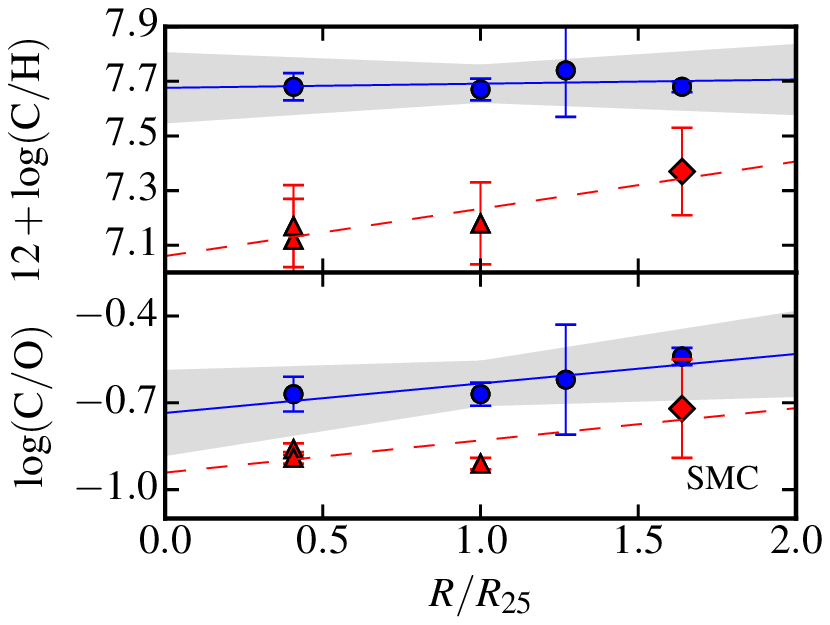}
 \caption{Spatial distribution of the C/H (upper panels) and C/O (lower panels) abundance ratios as a function of the fractional galactocentric distance ($R$/$R_{25}$) for the LMC (left-hand panels) and the SMC (right-hand panels). The C abundances have been determined from UV CELs, where red diamonds are taken from \citet{1995ApJ...443...64G} and red triangles from \citet{1982ApJ...252..461D}. Linear least-squares fits to these data are represented by the dashed red lines. Blue circles are our C/H and C/O ratios determined from RLs and their linear least-squares fits are represented by the blue continuous lines. The grey areas trace all the radial gradients compatible with the uncertainties of our least-squares linear fits computed through Monte Carlo simulations (see Section \ref{sec:radial_abun}).}
 \label{fig:cvo}
\end{figure*}

The results gathered in Table~\ref{tab:fits} indicate that the O/H gradients can be considered essentially flat in both galaxies. This is not a new result, \citet{1978MNRAS.184..569P} found a very small or flat radial gradient in the LMC -- of about $-$0.13 $\pm$ 0.09 $\mathrm{dex}\ (R/R_{\mathrm 25})^{-1}$ -- and a ``conspicuously absent'' gradient in the SMC. Several authors have studied the metallicity -- Fe/H ratio -- gradients of the MCs using different kinds of stars in order to investigate the formation mechanism of these galaxies. \citet{2009A&A...506.1137C} studied metallicities of asymptotic giant branch (AGB) stars of the LMC, finding a smooth gradient with a slope of  $-$0.20 $\pm$ 0.01 $\mathrm{dex}\ (R/R_{\mathrm 25})^{-1}$ up to 5 kpc. More recently, \citet{2016MNRAS.455.1855C} have confirmed a similar slope from observations of red giant brach (RGB) stars. However, using RR Lyrae variables, \citet{2010MNRAS.408L..76F} find a shallower gradient in metallicity for the LMC, with a slope between $-$0.06 and $-$0.04 $\mathrm{dex}\ (R/R_{\mathrm 25})^{-1}$ up and beyond a galactocentric radius of 5 kpc. In the case of the SMC, \citet[][]{2008AJ....136.1039C}, using the Ca\thinspace\textsc{ii} triplet in RGB stars found the first spectroscopic evidence of a metallicity gradient in this galaxy with the abundances of RGB stars related to an age gradient, with the youngest stars, which are the metal-richer ones, concentrated in the central regions of the galaxy. This result was supported by the study by \citet{2014MNRAS.442.1680D} which found an increasing fraction of young stars with decreasing galactocentric radius and a smooth radial metallicity gradient of 
$-$0.20 $\pm$ 0.03 $\mathrm{dex}\ (R/R_{\mathrm 25})^{-1}$ up to about 5 kpc. However, the study of the metallicity gradients in the SMC from AGB stars by \citet{2009A&A...506.1137C} obtains a negligible gradient. As we can see, there are contradictory results about the exact value of the slope of the metallicity radial gradients in both galaxies, however, all studies agree that, if such gradients exist, they should be certainly rather shallow.

The determination of C/H ratio in all our sample \hii\ regions allows us to study the spatial distribution of C/H and the C/O ratios -- and their radial gradients for the first time -- in the inner parts of the LMC and SMC. In Fig.~\ref{fig:cvo} we show the radial gradients of C/H (upper panels) and C/O (lower panels) in the LMC (left-hand panels) and the SMC (right-hand panels). In Table~\ref{tab:fits} we show the parameters of the fits plotted in Fig.~\ref{fig:cvo}. In contrast with the results that we have found in spiral galaxies, the slope of the C/H and C/O gradients in both galaxies can be considered practically zero. 

\begin{figure}
 \centering
  \includegraphics[]{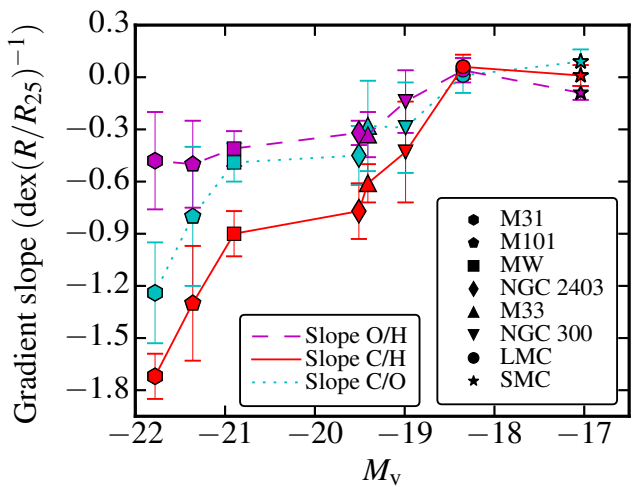}
  \caption{Slope of the O/H (magenta dashed line/symbols), C/H (red continuous line/symbols), and C/O (cian dotted line/symbols) radial gradients versus absolute magnitude, $M_{\mathrm{V}}$ of several spiral galaxies: SMC (stars), LMC (circles), NGC\,300 (down triangles), M33 (triangles), NGC\,2403 (diamonds), MW (squares), M101 (pentagons) and M31 (hexagons).}
 \label{fig:slope}
\end{figure}

\begin{figure*}
 \centering
  \includegraphics[width=0.49\textwidth]{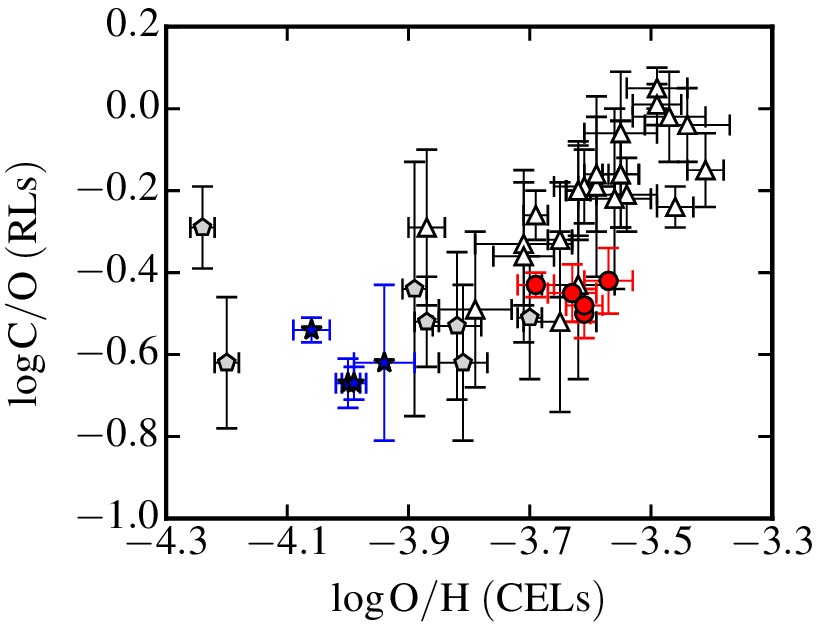}
  \includegraphics[width=0.49\textwidth]{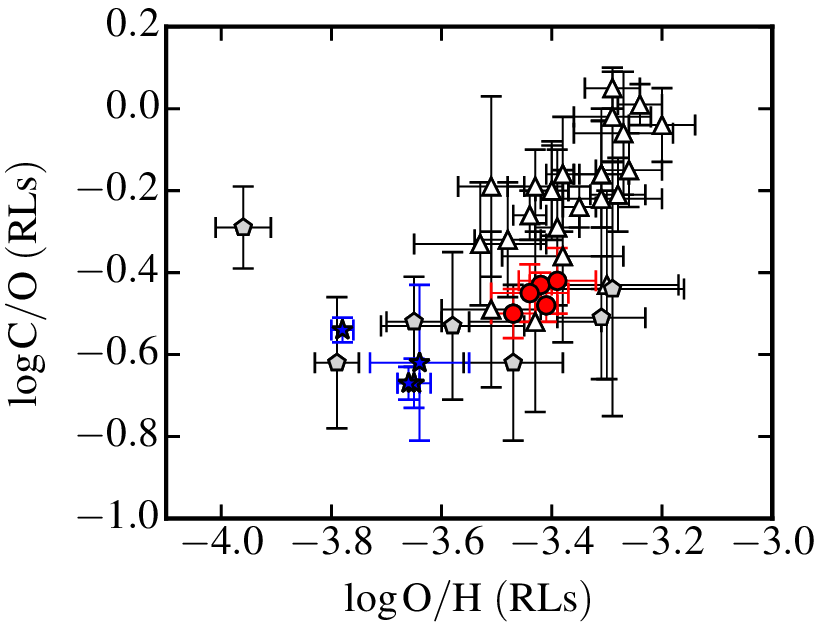}
 \caption{C/O \textit{versus} O/H ratios of \hii\ regions in different parent galaxies. Left panel: O/H ratios determined from CELs; right panel: O/H ratios determined from RLs. The red circles represent \hii\ regions of the LMC and the blue stars those of the SMC. The open triangles represent \hii\ regions of the Milky Way \citep{2005MNRAS.362..301G, 2006MNRAS.368..253G, 2007RMxAA..43....3G, 2004ApJS..153..501G, 2004MNRAS.355..229E, 2013MNRAS.433..382E} and in other galaxies \citep{2009ApJ...700..654E, 2016MNRAS.458.1866T}. The gray pentagons correspond to \hii\ regions in star-forming dwarf galaxies \citep[see][and references therein]{2014MNRAS.443..624E}. We recommend to add $\sim$0.1 dex to the values of log(O/H) to correct for dust depletion in the \hii\ regions. No correction is needed for their C/O ratios \citep[see][for details]{2014MNRAS.443..624E}.}
 \label{fig:evolution}
 \end{figure*}

In Fig.~\ref{fig:slope}, we have completed the information given in fig. 8 of \citet{2016MNRAS.458.1866T} -- which presented the slope of O/H, C/H and C/O gradients for different nearby spiral galaxies -- including the results we obtain for the MCs. This figure shows the slopes of the O/H, C/H and C/O radial gradients (magenta dashed line, red solid line and cyan dotted line, respectively) with respect to the absolute magnitude, $M_{\mathrm{V}}$, of the SMC, LMC, NGC\,300, M33, NGC\,2403, the Milky Way, M101 and M31. All the slopes represented in Fig.~\ref{fig:slope} have been calculated from RLs except the O/H ones of NGC\,2403 and M31 where the O gradient can only be determined from CELs. As we pointed out in Section~\ref{sec:radial_abun}, the slope of the O/H gradients seems to be rather independent of the kind of lines used to derive the abundances and, therefore, the use of the slope of O/H gradients based on CELs or RLs would give essentially the same results.

Fig.~\ref{fig:slope} reinforces the results obtained by \citet{2016MNRAS.458.1866T}. They presented the first indication that the slopes of the C/H -- and to a lesser extent C/O -- gradients show a clear correlation with the $M_{\mathrm{V}}$ of the galaxies. The more luminous galaxies --M31, M101 and the Milky Way-- show systematically steeper C/H and C/O gradients than the less luminous and less massive ones. All the spiral galaxies included in Fig.~\ref{fig:slope} show steeper gradients for 
C/H than for O/H, producing C/O negative gradients. In the case of the MCs -- the less massive galaxies and the only two irregular included in Fig.~\ref{fig:slope} -- the slopes of the gradients are about zero. According to an `inside-out' galaxy formation scenario, the negative C/O gradients are due to C enrichment the inner part of the galaxies \citep[see][]{2005ApJ...623..213C}. The low- and intermediate-mass stars have long timescales to inject C into the ISM. The newborn massive stars formed in more metal-rich environments, as the central parts of spiral galaxies, can further enrich the ISM via their stellar winds. Following this reasoning and the results of Fig.~\ref{fig:slope}, the mechanism of C enrichment would be more effective in the massive galaxies than the less massive ones.

Despite the fact that Fig.~\ref{fig:slope} seems to show a slight correlation between the slopes of the O/H gradient and the $M_{\mathrm{V}}$ of the galaxies, the uncertainties do not allow us to confirm such dependence. \citet{2015MNRAS.448.2030H} and \citet{2014A&A...563A..49S} studied the O/H gradients as a function of different properties of a large sample of spiral galaxies through O abundances determined by empirical calibrators. These authors found no correlation between the metallicity gradients and luminosity.

In Fig.~\ref{fig:evolution}, we show the behaviour of the C/O versus O/H abundance ratios determined from RLs. This is a new version of a plot presented in \citet{2014MNRAS.443..624E} 
and \citet{2016MNRAS.458.1866T}, which included data for the Milky Way and other spiral galaxies and star-forming dwarf galaxies but, in this case, adding our results for \hii\ regions of the MCs. We can see that the C/O ratios for the objects of the MCs have substantially smaller uncertainties than the majority of the objects included in the figure and fit perfectly the general shape for the C/O vs. O/H relation. The positions of the \hii\ regions of the SMC are consistent with the locus of giant {\hii} regions in star-forming dwarf galaxies. However, in the case of the points of the LMC, their positions are in between the star-forming dwarves and the most external {\hii} regions (around the isophotal radius, $R_{25}$) of the low-mass spiral galaxies NGC\,300 and M33 (see \citet{2016MNRAS.458.1866T}). This suggests that while the chemical evolution of the LMC seems to be similar to the external zones of small spiral galaxies, the SMC behaves as a typical star-forming dwarf galaxy.

\section{Conclusions}\label{sec:conclusion}

We present C and O abundances obtained from deep spectra of \hii\ regions in the Magellanic Clouds. The sample comprises 5 \hii\ regions in the LMC and 4 in the SMC. We calculated the electron temperature using direct determinations for each object. The C/H ratios have been determined from RLs of \cii\ and the O/H ratios from CELs and RLs of \oii. In addition, we have derived C/O ratios from RLs in all the objects of the sample.

The ADF(O$^{2+}$) values show a fairly small dispersion in both galaxies. Only N44C - the \hii\ region with the highest ionization degree - presents an ADF(O$^{2+}$) higher than the rest of the objects of the LMC. We have studied the behaviour of the ADF(O$^{2+}$) in \hii\ regions of several galaxies as a function of the total O abundances (from CELs and RLs). The results are complex to interpret. The ADF(O$^{2+}$) values show a minimum value at 12 + log(O/H) $\sim$ 8.4 or 8.5 and increase for higher and lower O abundances. The low dispersion of ADF(O$^{2+}$) in the low-metallicity objects allows us to suggest a metallicity dependence of the ADF. However, this trend doesn't seem so clear for high-metallicity objects -most of them Galactic \hii\ regions- due to the high dispersion of their ADF(O$^{2+}$) values.

We compare O and C nebular abundances with stellar ones for a sample of B-type stars of N11 in the LMC and N66 in the SMC. We find that the O and C abundances derived for B-type stars agree better with nebular ones derived from CELs. The comparison between stellar and nebular abundances in other galaxies provides contradictory results. With the purpose to understand these differences, we compare nebular and stellar abundances of a sample of star-forming regions in different galaxies. We find that the stellar abundances show better agreement with nebular abundances from CELs in low-metallicity regions and with RLs in high-metallicity regions. 

In agreement with previous works, we find that the radial gradients of O/H, C/H and C/O are almost flat in both galaxies. Our results for the MCs reinforce the trend outlined by \citet{2016MNRAS.458.1866T}, who found that the more luminous galaxies show systematically steeper C/H and C/O gradients than the less luminous ones. In addition, we studied the behaviour of C/O versus O/H in both galaxy. The results suggest that the chemical evolution of the LMC seems to be similar to the external zones of small spiral galaxies while the SMC behaves as a typical star-forming dwarf galaxy.

\section*{Acknowledgements}

This work is based on observations collected at the European Southern Observatory, Chile, proposal numbers ESO 092.C-0191(A) and ESO 60.A-9022(A). LTSC is supported by the FPI Program by the Ministerio de Econom\'ia y Competitividad (MINECO) under grant AYA2011-22614. This project project is also partially funded by MINECO under grant AYA2015-65205-P. GDG and MR acknowledge support from the Mexican CONACYT grant CB-2014-240562. AMD acknowledges support from the FONDECYT project 3140383.



\bibliographystyle{mnras}
\bibliography{bibtex}

\begin{thebibliography}{}
\makeatletter
\relax
\def\mn@urlcharsother{\let\do\@makeother \do\$\do\&\do\#\do\^\do\_\do\%\do\~}
\def\mn@doi{\begingroup\mn@urlcharsother \@ifnextchar [ {\mn@doi@}
  {\mn@doi@[]}}
\def\mn@doi@[#1]#2{\def\@tempa{#1}\ifx\@tempa\@empty \href
  {http://dx.doi.org/#2} {doi:#2}\else \href {http://dx.doi.org/#2} {#1}\fi
  \endgroup}
\def\mn@eprint#1#2{\mn@eprint@#1:#2::\@nil}
\def\mn@eprint@arXiv#1{\href {http://arxiv.org/abs/#1} {{\tt arXiv:#1}}}
\def\mn@eprint@dblp#1{\href {http://dblp.uni-trier.de/rec/bibtex/#1.xml}
  {dblp:#1}}
\def\mn@eprint@#1:#2:#3:#4\@nil{\def\@tempa {#1}\def\@tempb {#2}\def\@tempc
  {#3}\ifx \@tempc \@empty \let \@tempc \@tempb \let \@tempb \@tempa \fi \ifx
  \@tempb \@empty \def\@tempb {arXiv}\fi \@ifundefined
  {mn@eprint@\@tempb}{\@tempb:\@tempc}{\expandafter \expandafter \csname
  mn@eprint@\@tempb\endcsname \expandafter{\@tempc}}}

\bibitem[\protect\citeauthoryear{{Ballester}, {Modigliani}, {Boitquin},
  {Cristiani}, {Hanuschik}, {Kaufer}  \& {Wolf}}{{Ballester}
  et~al.}{2000}]{2000Msngr.101...31B}
{Ballester} P.,  {Modigliani} A.,  {Boitquin} O.,  {Cristiani} S.,  {Hanuschik}
  R.,  {Kaufer} A.,   {Wolf} S.,  2000, The Messenger, \href
  {http://adsabs.harvard.edu/abs/2000Msngr.101...31B} {101, 31}

\bibitem[\protect\citeauthoryear{{Berg}, {Skillman}, {Henry}, {Erb}  \&
  {Carigi}}{{Berg} et~al.}{2016}]{2016ApJ...827..126B}
{Berg} D.~A.,  {Skillman} E.~D.,  {Henry} R.~B.~C.,  {Erb} D.~K.,   {Carigi}
  L.,  2016, \mn@doi [\apj] {10.3847/0004-637X/827/2/126}, \href
  {http://adsabs.harvard.edu/abs/2016ApJ...827..126B} {827, 126}

\bibitem[\protect\citeauthoryear{{Bothun} \& {Thompson}}{{Bothun} \&
  {Thompson}}{1988}]{1988AJ.....96..877B}
{Bothun} G.~D.,  {Thompson} I.~B.,  1988, \mn@doi [\aj] {10.1086/114854}, \href
  {http://adsabs.harvard.edu/abs/1988AJ.....96..877B} {96, 877}

\bibitem[\protect\citeauthoryear{{Bresolin}, {Gieren}, {Kudritzki},
  {Pietrzy{\'n}ski}, {Urbaneja}  \& {Carraro}}{{Bresolin}
  et~al.}{2009}]{2009ApJ...700..309B}
{Bresolin} F.,  {Gieren} W.,  {Kudritzki} R.-P.,  {Pietrzy{\'n}ski} G.,
  {Urbaneja} M.~A.,   {Carraro} G.,  2009, \mn@doi [\apj]
  {10.1088/0004-637X/700/1/309}, \href
  {http://adsabs.harvard.edu/abs/2009ApJ...700..309B} {700, 309}

\bibitem[\protect\citeauthoryear{{Bresolin}, {Kudritzki}, {Urbaneja}, {Gieren},
  {Ho}  \& {Pietrzy{\'n}ski}}{{Bresolin} et~al.}{2016}]{2016ApJ...830...64B}
{Bresolin} F.,  {Kudritzki} R.-P.,  {Urbaneja} M.~A.,  {Gieren} W.,  {Ho}
  I.-T.,   {Pietrzy{\'n}ski} G.,  2016, \mn@doi [\apj]
  {10.3847/0004-637X/830/2/64}, \href
  {http://adsabs.harvard.edu/abs/2016ApJ...830...64B} {830, 64}

\bibitem[\protect\citeauthoryear{{Butler} \& {Zeippen}}{{Butler} \&
  {Zeippen}}{1989}]{1989AA...208..337B}
{Butler} K.,  {Zeippen} C.~J.,  1989, \aap, \href
  {http://adsabs.harvard.edu/abs/1989A%26A...208..337B} {208, 337}

\bibitem[\protect\citeauthoryear{{Carigi}, {Peimbert}, {Esteban}  \&
  {Garc{\'{\i}}a-Rojas}}{{Carigi} et~al.}{2005}]{2005ApJ...623..213C}
{Carigi} L.,  {Peimbert} M.,  {Esteban} C.,   {Garc{\'{\i}}a-Rojas} J.,  2005,
  \mn@doi [\apj] {10.1086/428491}, \href
  {http://adsabs.harvard.edu/abs/2005ApJ...623..213C} {623, 213}

\bibitem[\protect\citeauthoryear{{Carrera}, {Gallart}, {Aparicio}, {Costa},
  {M{\'e}ndez}  \& {No{\"e}l}}{{Carrera} et~al.}{2008}]{2008AJ....136.1039C}
{Carrera} R.,  {Gallart} C.,  {Aparicio} A.,  {Costa} E.,  {M{\'e}ndez} R.~A.,
   {No{\"e}l} N.~E.~D.,  2008, \mn@doi [\aj] {10.1088/0004-6256/136/3/1039},
  \href {http://adsabs.harvard.edu/abs/2008AJ....136.1039C} {136, 1039}

\bibitem[\protect\citeauthoryear{{Choudhury}, {Subramaniam}  \&
  {Cole}}{{Choudhury} et~al.}{2016}]{2016MNRAS.455.1855C}
{Choudhury} S.,  {Subramaniam} A.,   {Cole} A.~A.,  2016, \mn@doi [\mnras]
  {10.1093/mnras/stv2414}, \href
  {http://adsabs.harvard.edu/abs/2016MNRAS.455.1855C} {455, 1855}

\bibitem[\protect\citeauthoryear{{Cioni}}{{Cioni}}{2009}]{2009A&A...506.1137C}
{Cioni} M.-R.~L.,  2009, \mn@doi [\aap] {10.1051/0004-6361/200912138}, \href
  {http://adsabs.harvard.edu/abs/2009A%26A...506.1137C} {506, 1137}

\bibitem[\protect\citeauthoryear{{Davey}, {Storey}  \& {Kisielius}}{{Davey}
  et~al.}{2000}]{2000A&AS..142...85D}
{Davey} A.~R.,  {Storey} P.~J.,   {Kisielius} R.,  2000, \mn@doi [\aaps]
  {10.1051/aas:2000139}, \href
  {http://adsabs.harvard.edu/abs/2000A%26AS..142...85D} {142, 85}

\bibitem[\protect\citeauthoryear{{Delgado-Inglada}, {Morisset}  \&
  {Stasi{\'n}ska}}{{Delgado-Inglada} et~al.}{2014}]{2014MNRAS.440..536D}
{Delgado-Inglada} G.,  {Morisset} C.,   {Stasi{\'n}ska} G.,  2014, \mn@doi
  [\mnras] {10.1093/mnras/stu341}, \href
  {http://adsabs.harvard.edu/abs/2014MNRAS.440..536D} {440, 536}

\bibitem[\protect\citeauthoryear{{Dobbie}, {Cole}, {Subramaniam}  \&
  {Keller}}{{Dobbie} et~al.}{2014}]{2014MNRAS.442.1680D}
{Dobbie} P.~D.,  {Cole} A.~A.,  {Subramaniam} A.,   {Keller} S.,  2014, \mn@doi
  [\mnras] {10.1093/mnras/stu926}, \href
  {http://adsabs.harvard.edu/abs/2014MNRAS.442.1680D} {442, 1680}

\bibitem[\protect\citeauthoryear{{Dopita}, {Sutherland}, {Nicholls}, {Kewley}
  \& {Vogt}}{{Dopita} et~al.}{2013}]{2013ApJS..208...10D}
{Dopita} M.~A.,  {Sutherland} R.~S.,  {Nicholls} D.~C.,  {Kewley} L.~J.,
  {Vogt} F.~P.~A.,  2013, \mn@doi [\apjs] {10.1088/0067-0049/208/1/10}, \href
  {http://adsabs.harvard.edu/abs/2013ApJS..208...10D} {208, 10}

\bibitem[\protect\citeauthoryear{{Dufour}}{{Dufour}}{1975}]{1975ApJ...195..315D}
{Dufour} R.~J.,  1975, \mn@doi [\apj] {10.1086/153330}, \href
  {http://adsabs.harvard.edu/abs/1975ApJ...195..315D} {195, 315}

\bibitem[\protect\citeauthoryear{{Dufour}, {Shields}  \& {Talbot}}{{Dufour}
  et~al.}{1982}]{1982ApJ...252..461D}
{Dufour} R.~J.,  {Shields} G.~A.,   {Talbot} Jr. R.~J.,  1982, \mn@doi [\apj]
  {10.1086/159574}, \href {http://adsabs.harvard.edu/abs/1982ApJ...252..461D}
  {252, 461}

\bibitem[\protect\citeauthoryear{{Esteban}, {Peimbert}, {Torres-Peimbert}  \&
  {Escalante}}{{Esteban} et~al.}{1998}]{1998MNRAS.295..401E}
{Esteban} C.,  {Peimbert} M.,  {Torres-Peimbert} S.,   {Escalante} V.,  1998,
  \mn@doi [\mnras] {10.1046/j.1365-8711.1998.01335.x}, \href
  {http://adsabs.harvard.edu/abs/1998MNRAS.295..401E} {295, 401}

\bibitem[\protect\citeauthoryear{{Esteban}, {Peimbert}, {Torres-Peimbert}  \&
  {Rodr{\'{\i}}guez}}{{Esteban} et~al.}{2002}]{2002ApJ...581..241E}
{Esteban} C.,  {Peimbert} M.,  {Torres-Peimbert} S.,   {Rodr{\'{\i}}guez} M.,
  2002, \mn@doi [\apj] {10.1086/344104}, \href
  {http://adsabs.harvard.edu/abs/2002ApJ...581..241E} {581, 241}

\bibitem[\protect\citeauthoryear{{Esteban}, {Peimbert}, {Garc{\'{\i}}a-Rojas},
  {Ruiz}, {Peimbert}  \& {Rodr{\'{\i}}guez}}{{Esteban}
  et~al.}{2004}]{2004MNRAS.355..229E}
{Esteban} C.,  {Peimbert} M.,  {Garc{\'{\i}}a-Rojas} J.,  {Ruiz} M.~T.,
  {Peimbert} A.,   {Rodr{\'{\i}}guez} M.,  2004, \mn@doi [\mnras]
  {10.1111/j.1365-2966.2004.08313.x}, \href
  {http://adsabs.harvard.edu/abs/2004MNRAS.355..229E} {355, 229}

\bibitem[\protect\citeauthoryear{{Esteban}, {Garc{\'{\i}}a-Rojas}, {Peimbert},
  {Peimbert}, {Ruiz}, {Rodr{\'{\i}}guez}  \& {Carigi}}{{Esteban}
  et~al.}{2005}]{2005ApJ...618L..95E}
{Esteban} C.,  {Garc{\'{\i}}a-Rojas} J.,  {Peimbert} M.,  {Peimbert} A.,
  {Ruiz} M.~T.,  {Rodr{\'{\i}}guez} M.,   {Carigi} L.,  2005, \mn@doi [\apjl]
  {10.1086/426889}, \href {http://adsabs.harvard.edu/abs/2005ApJ...618L..95E}
  {618, L95}

\bibitem[\protect\citeauthoryear{{Esteban}, {Bresolin}, {Peimbert},
  {Garc{\'{\i}}a-Rojas}, {Peimbert}  \& {Mesa-Delgado}}{{Esteban}
  et~al.}{2009}]{2009ApJ...700..654E}
{Esteban} C.,  {Bresolin} F.,  {Peimbert} M.,  {Garc{\'{\i}}a-Rojas} J.,
  {Peimbert} A.,   {Mesa-Delgado} A.,  2009, \mn@doi [\apj]
  {10.1088/0004-637X/700/1/654}, \href
  {http://adsabs.harvard.edu/abs/2009ApJ...700..654E} {700, 654}

\bibitem[\protect\citeauthoryear{{Esteban}, {Carigi}, {Copetti},
  {Garc{\'{\i}}a-Rojas}, {Mesa-Delgado}, {Casta{\~n}eda}  \&
  {P{\'e}quignot}}{{Esteban} et~al.}{2013}]{2013MNRAS.433..382E}
{Esteban} C.,  {Carigi} L.,  {Copetti} M.~V.~F.,  {Garc{\'{\i}}a-Rojas} J.,
  {Mesa-Delgado} A.,  {Casta{\~n}eda} H.~O.,   {P{\'e}quignot} D.,  2013,
  \mn@doi [\mnras] {10.1093/mnras/stt730}, \href
  {http://adsabs.harvard.edu/abs/2013MNRAS.433..382E} {433, 382}

\bibitem[\protect\citeauthoryear{{Esteban}, {Garc{\'{\i}}a-Rojas}, {Carigi},
  {Peimbert}, {Bresolin}, {L{\'o}pez-S{\'a}nchez}  \& {Mesa-Delgado}}{{Esteban}
  et~al.}{2014}]{2014MNRAS.443..624E}
{Esteban} C.,  {Garc{\'{\i}}a-Rojas} J.,  {Carigi} L.,  {Peimbert} M.,
  {Bresolin} F.,  {L{\'o}pez-S{\'a}nchez} A.~R.,   {Mesa-Delgado} A.,  2014,
  \mn@doi [\mnras] {10.1093/mnras/stu1177}, \href
  {http://adsabs.harvard.edu/abs/2014MNRAS.443..624E} {443, 624}

\bibitem[\protect\citeauthoryear{{Esteban}, {Mesa-Delgado}, {Morisset}  \&
  {Garc{\'{\i}}a-Rojas}}{{Esteban} et~al.}{2016}]{2016MNRAS.460.4038E}
{Esteban} C.,  {Mesa-Delgado} A.,  {Morisset} C.,   {Garc{\'{\i}}a-Rojas} J.,
  2016, \mn@doi [\mnras] {10.1093/mnras/stw1243}, \href
  {http://adsabs.harvard.edu/abs/2016MNRAS.460.4038E} {460, 4038}

\bibitem[\protect\citeauthoryear{{Feast}, {Abedigamba}  \& {Whitelock}}{{Feast}
  et~al.}{2010}]{2010MNRAS.408L..76F}
{Feast} M.~W.,  {Abedigamba} O.~P.,   {Whitelock} P.~A.,  2010, \mn@doi
  [\mnras] {10.1111/j.1745-3933.2010.00933.x}, \href
  {http://adsabs.harvard.edu/abs/2010MNRAS.408L..76F} {408, L76}

\bibitem[\protect\citeauthoryear{{Froese Fischer} \& {Tachiev}}{{Froese
  Fischer} \& {Tachiev}}{2004}]{2004ADNDT..87....1F}
{Froese Fischer} C.,  {Tachiev} G.,  2004, \mn@doi [Atomic Data and Nuclear
  Data Tables] {10.1016/j.adt.2004.02.001}, \href
  {http://adsabs.harvard.edu/abs/2004ADNDT..87....1F} {87, 1}

\bibitem[\protect\citeauthoryear{{Garc{\'{\i}}a-Rojas} \&
  {Esteban}}{{Garc{\'{\i}}a-Rojas} \& {Esteban}}{2007}]{2007ApJ...670..457G}
{Garc{\'{\i}}a-Rojas} J.,  {Esteban} C.,  2007, \mn@doi [\apj]
  {10.1086/521871}, \href {http://adsabs.harvard.edu/abs/2007ApJ...670..457G}
  {670, 457}

\bibitem[\protect\citeauthoryear{{Garc{\'{\i}}a-Rojas}, {Esteban}, {Peimbert},
  {Rodr{\'{\i}}guez}, {Ruiz}  \& {Peimbert}}{{Garc{\'{\i}}a-Rojas}
  et~al.}{2004}]{2004ApJS..153..501G}
{Garc{\'{\i}}a-Rojas} J.,  {Esteban} C.,  {Peimbert} M.,  {Rodr{\'{\i}}guez}
  M.,  {Ruiz} M.~T.,   {Peimbert} A.,  2004, \mn@doi [\apjs] {10.1086/421909},
  \href {http://adsabs.harvard.edu/abs/2004ApJS..153..501G} {153, 501}

\bibitem[\protect\citeauthoryear{{Garc{\'{\i}}a-Rojas}, {Esteban}, {Peimbert},
  {Peimbert}, {Rodr{\'{\i}}guez}  \& {Ruiz}}{{Garc{\'{\i}}a-Rojas}
  et~al.}{2005}]{2005MNRAS.362..301G}
{Garc{\'{\i}}a-Rojas} J.,  {Esteban} C.,  {Peimbert} A.,  {Peimbert} M.,
  {Rodr{\'{\i}}guez} M.,   {Ruiz} M.~T.,  2005, \mn@doi [\mnras]
  {10.1111/j.1365-2966.2005.09302.x}, \href
  {http://adsabs.harvard.edu/abs/2005MNRAS.362..301G} {362, 301}

\bibitem[\protect\citeauthoryear{{Garc{\'{\i}}a-Rojas}, {Esteban}, {Peimbert},
  {Costado}, {Rodr{\'{\i}}guez}, {Peimbert}  \& {Ruiz}}{{Garc{\'{\i}}a-Rojas}
  et~al.}{2006}]{2006MNRAS.368..253G}
{Garc{\'{\i}}a-Rojas} J.,  {Esteban} C.,  {Peimbert} M.,  {Costado} M.~T.,
  {Rodr{\'{\i}}guez} M.,  {Peimbert} A.,   {Ruiz} M.~T.,  2006, \mn@doi
  [\mnras] {10.1111/j.1365-2966.2006.10105.x}, \href
  {http://adsabs.harvard.edu/abs/2006MNRAS.368..253G} {368, 253}

\bibitem[\protect\citeauthoryear{{Garc{\'{\i}}a-Rojas}, {Esteban}, {Peimbert},
  {Rodr{\'{\i}}guez}, {Peimbert}  \& {Ruiz}}{{Garc{\'{\i}}a-Rojas}
  et~al.}{2007}]{2007RMxAA..43....3G}
{Garc{\'{\i}}a-Rojas} J.,  {Esteban} C.,  {Peimbert} A.,  {Rodr{\'{\i}}guez}
  M.,  {Peimbert} M.,   {Ruiz} M.~T.,  2007, \rmxaa, \href
  {http://adsabs.harvard.edu/abs/2007RMxAA..43....3G} {43, 3}

\bibitem[\protect\citeauthoryear{{Garc{\'{\i}}a-Rojas}, {Sim{\'o}n-D{\'{\i}}az}
   \& {Esteban}}{{Garc{\'{\i}}a-Rojas} et~al.}{2014}]{2014A&A...571A..93G}
{Garc{\'{\i}}a-Rojas} J.,  {Sim{\'o}n-D{\'{\i}}az} S.,   {Esteban} C.,  2014,
  \mn@doi [\aap] {10.1051/0004-6361/201424660}, \href
  {http://adsabs.harvard.edu/abs/2014A%26A...571A..93G} {571, A93}

\bibitem[\protect\citeauthoryear{{Garc{\'{\i}}a}, {Herrero}, {Najarro},
  {Lennon}  \& {Alejandro Urbaneja}}{{Garc{\'{\i}}a}
  et~al.}{2014}]{2014ApJ...788...64G}
{Garc{\'{\i}}a} M.,  {Herrero} A.,  {Najarro} F.,  {Lennon} D.~J.,   {Alejandro
  Urbaneja} M.,  2014, \mn@doi [\apj] {10.1088/0004-637X/788/1/64}, \href
  {http://adsabs.harvard.edu/abs/2014ApJ...788...64G} {788, 64}

\bibitem[\protect\citeauthoryear{{Garnett}, {Skillman}, {Dufour}, {Peimbert},
  {Torres-Peimbert}, {Terlevich}, {Terlevich}  \& {Shields}}{{Garnett}
  et~al.}{1995}]{1995ApJ...443...64G}
{Garnett} D.~R.,  {Skillman} E.~D.,  {Dufour} R.~J.,  {Peimbert} M.,
  {Torres-Peimbert} S.,  {Terlevich} R.,  {Terlevich} E.,   {Shields} G.~A.,
  1995, \mn@doi [\apj] {10.1086/175503}, \href
  {http://adsabs.harvard.edu/abs/1995ApJ...443...64G} {443, 64}

\bibitem[\protect\citeauthoryear{{Garnett}, {Shields}, {Peimbert},
  {Torres-Peimbert}, {Skillman}, {Dufour}, {Terlevich}  \&
  {Terlevich}}{{Garnett} et~al.}{1999}]{1999ApJ...513..168G}
{Garnett} D.~R.,  {Shields} G.~A.,  {Peimbert} M.,  {Torres-Peimbert} S.,
  {Skillman} E.~D.,  {Dufour} R.~J.,  {Terlevich} E.,   {Terlevich} R.~J.,
  1999, \mn@doi [\apj] {10.1086/306860}, \href
  {http://adsabs.harvard.edu/abs/1999ApJ...513..168G} {513, 168}

\bibitem[\protect\citeauthoryear{{Groenewegen}}{{Groenewegen}}{2000}]{2000A&A...363..901G}
{Groenewegen} M.~A.~T.,  2000, \aap, \href
  {http://adsabs.harvard.edu/abs/2000A%26A...363..901G} {363, 901}

\bibitem[\protect\citeauthoryear{{Henize}}{{Henize}}{1956}]{1956ApJS....2..315H}
{Henize} K.~G.,  1956, \mn@doi [\apjs] {10.1086/190025}, \href
  {http://adsabs.harvard.edu/abs/1956ApJS....2..315H} {2, 315}

\bibitem[\protect\citeauthoryear{{Ho}, {Kudritzki}, {Kewley}, {Zahid},
  {Dopita}, {Bresolin}  \& {Rupke}}{{Ho} et~al.}{2015}]{2015MNRAS.448.2030H}
{Ho} I.-T.,  {Kudritzki} R.-P.,  {Kewley} L.~J.,  {Zahid} H.~J.,  {Dopita}
  M.~A.,  {Bresolin} F.,   {Rupke} D.~S.~N.,  2015, \mn@doi [\mnras]
  {10.1093/mnras/stv067}, \href
  {http://adsabs.harvard.edu/abs/2015MNRAS.448.2030H} {448, 2030}

\bibitem[\protect\citeauthoryear{{Howarth}}{{Howarth}}{1983}]{1983MNRAS.203..301H}
{Howarth} I.~D.,  1983, \mn@doi [\mnras] {10.1093/mnras/203.2.301}, \href
  {http://adsabs.harvard.edu/abs/1983MNRAS.203..301H} {203, 301}

\bibitem[\protect\citeauthoryear{{Hunter} et~al.,}{{Hunter}
  et~al.}{2009}]{2009A&A...496..841H}
{Hunter} I.,  et~al., 2009, \mn@doi [\aap] {10.1051/0004-6361/200809925}, \href
  {http://adsabs.harvard.edu/abs/2009A%26A...496..841H} {496, 841}

\bibitem[\protect\citeauthoryear{{Kennicutt}, {Bresolin}, {Bomans}, {Bothun}
  \& {Thompson}}{{Kennicutt} et~al.}{1995}]{1995AJ....109..594K}
{Kennicutt} Jr. R.~C.,  {Bresolin} F.,  {Bomans} D.~J.,  {Bothun} G.~D.,
  {Thompson} I.~B.,  1995, \mn@doi [\aj] {10.1086/117304}, \href
  {http://adsabs.harvard.edu/abs/1995AJ....109..594K} {109, 594}

\bibitem[\protect\citeauthoryear{{Kisielius}, {Storey}, {Ferland}  \&
  {Keenan}}{{Kisielius} et~al.}{2009}]{2009MNRAS.397..903K}
{Kisielius} R.,  {Storey} P.~J.,  {Ferland} G.~J.,   {Keenan} F.~P.,  2009,
  \mn@doi [\mnras] {10.1111/j.1365-2966.2009.14989.x}, \href
  {http://adsabs.harvard.edu/abs/2009MNRAS.397..903K} {397, 903}

\bibitem[\protect\citeauthoryear{{Liu}, {Storey}, {Barlow}, {Danziger}, {Cohen}
   \& {Bryce}}{{Liu} et~al.}{2000}]{2000MNRAS.312..585L}
{Liu} X.-W.,  {Storey} P.~J.,  {Barlow} M.~J.,  {Danziger} I.~J.,  {Cohen} M.,
   {Bryce} M.,  2000, \mn@doi [\mnras] {10.1046/j.1365-8711.2000.03167.x},
  \href {http://adsabs.harvard.edu/abs/2000MNRAS.312..585L} {312, 585}

\bibitem[\protect\citeauthoryear{{L{\'o}pez-S{\'a}nchez}, {Esteban}  \&
  {Garc{\'{\i}}a-Rojas}}{{L{\'o}pez-S{\'a}nchez}
  et~al.}{2006}]{2006A&A...449..997L}
{L{\'o}pez-S{\'a}nchez} {\'A}.~R.,  {Esteban} C.,   {Garc{\'{\i}}a-Rojas} J.,
  2006, \mn@doi [\aap] {10.1051/0004-6361:20053119}, \href
  {http://adsabs.harvard.edu/abs/2006A%26A...449..997L} {449, 997}

\bibitem[\protect\citeauthoryear{{L{\'o}pez-S{\'a}nchez}, {Esteban},
  {Garc{\'{\i}}a-Rojas}, {Peimbert}  \&
  {Rodr{\'{\i}}guez}}{{L{\'o}pez-S{\'a}nchez}
  et~al.}{2007}]{2007ApJ...656..168L}
{L{\'o}pez-S{\'a}nchez} {\'A}.~R.,  {Esteban} C.,  {Garc{\'{\i}}a-Rojas} J.,
  {Peimbert} M.,   {Rodr{\'{\i}}guez} M.,  2007, \mn@doi [\apj]
  {10.1086/510112}, \href {http://adsabs.harvard.edu/abs/2007ApJ...656..168L}
  {656, 168}

\bibitem[\protect\citeauthoryear{{Luridiana}, {Morisset}  \&
  {Shaw}}{{Luridiana} et~al.}{2015}]{2015A&A...573A..42L}
{Luridiana} V.,  {Morisset} C.,   {Shaw} R.~A.,  2015, \mn@doi [\aap]
  {10.1051/0004-6361/201323152}, \href
  {http://adsabs.harvard.edu/abs/2015A%26A...573A..42L} {573, A42}

\bibitem[\protect\citeauthoryear{{Mathis}, {Torres-Peimbert}  \&
  {Peimbert}}{{Mathis} et~al.}{1998}]{1998ApJ...495..328M}
{Mathis} J.~S.,  {Torres-Peimbert} S.,   {Peimbert} M.,  1998, \mn@doi [\apj]
  {10.1086/305254}, \href {http://adsabs.harvard.edu/abs/1998ApJ...495..328M}
  {495, 328}

\bibitem[\protect\citeauthoryear{{Mendoza}}{{Mendoza}}{1983}]{1983IAUS..103..143M}
{Mendoza} C.,  1983, in {Flower} D.~R.,  ed.,  IAU Symposium Vol. 103,
  Planetary Nebulae. pp 143--172

\bibitem[\protect\citeauthoryear{{Mesa-Delgado}, {Esteban}  \&
  {Garc{\'{\i}}a-Rojas}}{{Mesa-Delgado} et~al.}{2008}]{2008ApJ...675..389M}
{Mesa-Delgado} A.,  {Esteban} C.,   {Garc{\'{\i}}a-Rojas} J.,  2008, \mn@doi
  [\apj] {10.1086/524296}, \href
  {http://adsabs.harvard.edu/abs/2008ApJ...675..389M} {675, 389}

\bibitem[\protect\citeauthoryear{{Mesa-Delgado}, {Esteban},
  {Garc{\'{\i}}a-Rojas}, {Luridiana}, {Bautista}, {Rodr{\'{\i}}guez},
  {L{\'o}pez-Mart{\'{\i}}n}  \& {Peimbert}}{{Mesa-Delgado}
  et~al.}{2009}]{2009MNRAS.395..855M}
{Mesa-Delgado} A.,  {Esteban} C.,  {Garc{\'{\i}}a-Rojas} J.,  {Luridiana} V.,
  {Bautista} M.,  {Rodr{\'{\i}}guez} M.,  {L{\'o}pez-Mart{\'{\i}}n} L.,
  {Peimbert} M.,  2009, \mn@doi [\mnras] {10.1111/j.1365-2966.2009.14554.x},
  \href {http://adsabs.harvard.edu/abs/2009MNRAS.395..855M} {395, 855}

\bibitem[\protect\citeauthoryear{{Mesa-Delgado}, {N{\'u}{\~n}ez-D{\'{\i}}az},
  {Esteban}, {L{\'o}pez-Mart{\'{\i}}n}  \&
  {Garc{\'{\i}}a-Rojas}}{{Mesa-Delgado} et~al.}{2011}]{2011MNRAS.417..420M}
{Mesa-Delgado} A.,  {N{\'u}{\~n}ez-D{\'{\i}}az} M.,  {Esteban} C.,
  {L{\'o}pez-Mart{\'{\i}}n} L.,   {Garc{\'{\i}}a-Rojas} J.,  2011, \mn@doi
  [\mnras] {10.1111/j.1365-2966.2011.19278.x}, \href
  {http://adsabs.harvard.edu/abs/2011MNRAS.417..420M} {417, 420}

\bibitem[\protect\citeauthoryear{{Mesa-Delgado}, {Esteban},
  {Garc{\'{\i}}a-Rojas}, {Reyes-P{\'e}rez}, {Morisset}  \&
  {Bresolin}}{{Mesa-Delgado} et~al.}{2014}]{2014ApJ...785..100M}
{Mesa-Delgado} A.,  {Esteban} C.,  {Garc{\'{\i}}a-Rojas} J.,  {Reyes-P{\'e}rez}
  J.,  {Morisset} C.,   {Bresolin} F.,  2014, \mn@doi [\apj]
  {10.1088/0004-637X/785/2/100}, \href
  {http://adsabs.harvard.edu/abs/2014ApJ...785..100M} {785, 100}

\bibitem[\protect\citeauthoryear{{Nicholls}, {Dopita}  \&
  {Sutherland}}{{Nicholls} et~al.}{2012}]{2012ApJ...752..148N}
{Nicholls} D.~C.,  {Dopita} M.~A.,   {Sutherland} R.~S.,  2012, \mn@doi [\apj]
  {10.1088/0004-637X/752/2/148}, \href
  {http://adsabs.harvard.edu/abs/2012ApJ...752..148N} {752, 148}

\bibitem[\protect\citeauthoryear{{Nieva} \& {Przybilla}}{{Nieva} \&
  {Przybilla}}{2012}]{2012A&A...539A.143N}
{Nieva} M.-F.,  {Przybilla} N.,  2012, \mn@doi [\aap]
  {10.1051/0004-6361/201118158}, \href
  {http://adsabs.harvard.edu/abs/2012A%26A...539A.143N} {539, A143}

\bibitem[\protect\citeauthoryear{{Pagel}, {Edmunds}, {Fosbury}  \&
  {Webster}}{{Pagel} et~al.}{1978}]{1978MNRAS.184..569P}
{Pagel} B.~E.~J.,  {Edmunds} M.~G.,  {Fosbury} R.~A.~E.,   {Webster} B.~L.,
  1978, \mn@doi [\mnras] {10.1093/mnras/184.3.569}, \href
  {http://adsabs.harvard.edu/abs/1978MNRAS.184..569P} {184, 569}

\bibitem[\protect\citeauthoryear{{Parker} et~al.,}{{Parker}
  et~al.}{1998}]{1998AJ....116..180P}
{Parker} J.~W.,  et~al., 1998, \mn@doi [\aj] {10.1086/300419}, \href
  {http://adsabs.harvard.edu/abs/1998AJ....116..180P} {116, 180}

\bibitem[\protect\citeauthoryear{{Patrick}, {Evans}, {Davies}, {Kudritzki},
  {Gazak}, {Bergemann}, {Plez}  \& {Ferguson}}{{Patrick}
  et~al.}{2015}]{2015ApJ...803...14P}
{Patrick} L.~R.,  {Evans} C.~J.,  {Davies} B.,  {Kudritzki} R.-P.,  {Gazak}
  J.~Z.,  {Bergemann} M.,  {Plez} B.,   {Ferguson} A.~M.~N.,  2015, \mn@doi
  [\apj] {10.1088/0004-637X/803/1/14}, \href
  {http://adsabs.harvard.edu/abs/2015ApJ...803...14P} {803, 14}

\bibitem[\protect\citeauthoryear{{Pe{\~n}a-Guerrero}, {Peimbert}, {Peimbert}
  \& {Ruiz}}{{Pe{\~n}a-Guerrero} et~al.}{2012}]{2012ApJ...746..115P}
{Pe{\~n}a-Guerrero} M.~A.,  {Peimbert} A.,  {Peimbert} M.,   {Ruiz} M.~T.,
  2012, \mn@doi [\apj] {10.1088/0004-637X/746/2/115}, \href
  {http://adsabs.harvard.edu/abs/2012ApJ...746..115P} {746, 115}

\bibitem[\protect\citeauthoryear{{Peimbert}}{{Peimbert}}{2003}]{2003ApJ...584..735P}
{Peimbert} A.,  2003, \mn@doi [\apj] {10.1086/345793}, \href
  {http://adsabs.harvard.edu/abs/2003ApJ...584..735P} {584, 735}

\bibitem[\protect\citeauthoryear{{Peimbert} \& {Costero}}{{Peimbert} \&
  {Costero}}{1969}]{1969BOTT....5....3P}
{Peimbert} M.,  {Costero} R.,  1969, Boletin de los Observatorios Tonantzintla
  y Tacubaya, \href {http://adsabs.harvard.edu/abs/1969BOTT....5....3P} {5, 3}

\bibitem[\protect\citeauthoryear{{Peimbert} \& {Peimbert}}{{Peimbert} \&
  {Peimbert}}{2005}]{2005RMxAC..23....9P}
{Peimbert} A.,  {Peimbert} M.,  2005, in Rev. Mex. Astron. Astrofis. (Ser.
  Conf.). pp 9--14

\bibitem[\protect\citeauthoryear{{Peimbert} \& {Peimbert}}{{Peimbert} \&
  {Peimbert}}{2010}]{2010ApJ...724..791P}
{Peimbert} A.,  {Peimbert} M.,  2010, \mn@doi [\apj]
  {10.1088/0004-637X/724/1/791}, \href
  {http://adsabs.harvard.edu/abs/2010ApJ...724..791P} {724, 791}

\bibitem[\protect\citeauthoryear{{Peimbert} \& {Torres-Peimbert}}{{Peimbert} \&
  {Torres-Peimbert}}{1974}]{1974ApJ...193..327P}
{Peimbert} M.,  {Torres-Peimbert} S.,  1974, \mn@doi [\apj] {10.1086/153166},
  \href {http://adsabs.harvard.edu/abs/1974ApJ...193..327P} {193, 327}

\bibitem[\protect\citeauthoryear{{Peimbert} \& {Torres-Peimbert}}{{Peimbert} \&
  {Torres-Peimbert}}{1976}]{1976ApJ...203..581P}
{Peimbert} M.,  {Torres-Peimbert} S.,  1976, \mn@doi [\apj] {10.1086/154114},
  \href {http://adsabs.harvard.edu/abs/1976ApJ...203..581P} {203, 581}

\bibitem[\protect\citeauthoryear{{Peimbert}, {Pena}  \&
  {Torres-Peimbert}}{{Peimbert} et~al.}{1986}]{1986A&A...158..266P}
{Peimbert} M.,  {Pena} M.,   {Torres-Peimbert} S.,  1986, \aap, \href
  {http://adsabs.harvard.edu/abs/1986A%26A...158..266P} {158, 266}

\bibitem[\protect\citeauthoryear{{Peimbert}, {Peimbert}  \& {Ruiz}}{{Peimbert}
  et~al.}{2000}]{2000ApJ...541..688P}
{Peimbert} M.,  {Peimbert} A.,   {Ruiz} M.~T.,  2000, \mn@doi [\apj]
  {10.1086/309485}, \href {http://adsabs.harvard.edu/abs/2000ApJ...541..688P}
  {541, 688}

\bibitem[\protect\citeauthoryear{{Peimbert}, {Peimbert}  \& {Ruiz}}{{Peimbert}
  et~al.}{2005}]{2005ApJ...634.1056P}
{Peimbert} A.,  {Peimbert} M.,   {Ruiz} M.~T.,  2005, \mn@doi [\apj]
  {10.1086/444557}, \href {http://adsabs.harvard.edu/abs/2005ApJ...634.1056P}
  {634, 1056}

\bibitem[\protect\citeauthoryear{{Pietrzy{\'n}ski} et~al.,}{{Pietrzy{\'n}ski}
  et~al.}{2013}]{2013Natur.495...76P}
{Pietrzy{\'n}ski} G.,  et~al., 2013, \mn@doi [\nat] {10.1038/nature11878},
  \href {http://adsabs.harvard.edu/abs/2013Natur.495...76P} {495, 76}

\bibitem[\protect\citeauthoryear{{Podobedova}, {Kelleher}  \&
  {Wiese}}{{Podobedova} et~al.}{2009}]{2009JPCRD..38..171P}
{Podobedova} L.~I.,  {Kelleher} D.~E.,   {Wiese} W.~L.,  2009, \mn@doi [Journal
  of Physical and Chemical Reference Data] {10.1063/1.3032939}, \href
  {http://adsabs.harvard.edu/abs/2009JPCRD..38..171P} {38, 171}

\bibitem[\protect\citeauthoryear{{Porter}, {Ferland}, {Storey}  \&
  {Detisch}}{{Porter} et~al.}{2012}]{2012MNRAS.425L..28P}
{Porter} R.~L.,  {Ferland} G.~J.,  {Storey} P.~J.,   {Detisch} M.~J.,  2012,
  \mn@doi [\mnras] {10.1111/j.1745-3933.2012.01300.x}, \href
  {http://adsabs.harvard.edu/abs/2012MNRAS.425L..28P} {425, L28}

\bibitem[\protect\citeauthoryear{{Porter}, {Ferland}, {Storey}  \&
  {Detisch}}{{Porter} et~al.}{2013}]{2013MNRAS.433L..89P}
{Porter} R.~L.,  {Ferland} G.~J.,  {Storey} P.~J.,   {Detisch} M.~J.,  2013,
  \mn@doi [\mnras] {10.1093/mnrasl/slt049}, \href
  {http://adsabs.harvard.edu/abs/2013MNRAS.433L..89P} {433, L89}

\bibitem[\protect\citeauthoryear{{Ruiz}, {Peimbert}, {Peimbert}  \&
  {Esteban}}{{Ruiz} et~al.}{2003}]{2003ApJ...595..247R}
{Ruiz} M.~T.,  {Peimbert} A.,  {Peimbert} M.,   {Esteban} C.,  2003, \mn@doi
  [\apj] {10.1086/377255}, \href
  {http://adsabs.harvard.edu/abs/2003ApJ...595..247R} {595, 247}

\bibitem[\protect\citeauthoryear{{S{\'a}nchez} et~al.,}{{S{\'a}nchez}
  et~al.}{2014}]{2014A&A...563A..49S}
{S{\'a}nchez} S.~F.,  et~al., 2014, \mn@doi [\aap]
  {10.1051/0004-6361/201322343}, \href
  {http://adsabs.harvard.edu/abs/2014A%26A...563A..49S} {563, A49}

\bibitem[\protect\citeauthoryear{{Scowcroft}, {Freedman}, {Madore}, {Monson},
  {Persson}, {Rich}, {Seibert}  \& {Rigby}}{{Scowcroft}
  et~al.}{2016}]{2016ApJ...816...49S}
{Scowcroft} V.,  {Freedman} W.~L.,  {Madore} B.~F.,  {Monson} A.,  {Persson}
  S.~E.,  {Rich} J.,  {Seibert} M.,   {Rigby} J.~R.,  2016, \mn@doi [\apj]
  {10.3847/0004-637X/816/2/49}, \href
  {http://adsabs.harvard.edu/abs/2016ApJ...816...49S} {816, 49}

\bibitem[\protect\citeauthoryear{{Selier} \& {Heydari-Malayeri}}{{Selier} \&
  {Heydari-Malayeri}}{2012}]{2012A&A...545A..29S}
{Selier} R.,  {Heydari-Malayeri} M.,  2012, \mn@doi [\aap]
  {10.1051/0004-6361/201219706}, \href
  {http://adsabs.harvard.edu/abs/2012A%26A...545A..29S} {545, A29}

\bibitem[\protect\citeauthoryear{{Sim{\'o}n-D{\'{\i}}az} \&
  {Stasi{\'n}ska}}{{Sim{\'o}n-D{\'{\i}}az} \&
  {Stasi{\'n}ska}}{2011}]{2011A&A...526A..48S}
{Sim{\'o}n-D{\'{\i}}az} S.,  {Stasi{\'n}ska} G.,  2011, \mn@doi [\aap]
  {10.1051/0004-6361/201015512}, \href
  {http://adsabs.harvard.edu/abs/2011A%26A...526A..48S} {526, A48}

\bibitem[\protect\citeauthoryear{{Sim{\'o}n-D{\'{\i}}az}, {Nieva}, {Przybilla}
  \& {Stasi{\'n}ska}}{{Sim{\'o}n-D{\'{\i}}az}
  et~al.}{2011}]{2011BSRSL..80..255S}
{Sim{\'o}n-D{\'{\i}}az} S.,  {Nieva} M.~F.,  {Przybilla} N.,   {Stasi{\'n}ska}
  G.,  2011, Bulletin de la Societe Royale des Sciences de Liege, \href
  {http://adsabs.harvard.edu/abs/2011BSRSL..80..255S} {80, 255}

\bibitem[\protect\citeauthoryear{{Smartt}, {Crowther}, {Dufton}, {Lennon},
  {Kudritzki}, {Herrero}, {McCarthy}  \& {Bresolin}}{{Smartt}
  et~al.}{2001}]{2001MNRAS.325..257S}
{Smartt} S.~J.,  {Crowther} P.~A.,  {Dufton} P.~L.,  {Lennon} D.~J.,
  {Kudritzki} R.~P.,  {Herrero} A.,  {McCarthy} J.~K.,   {Bresolin} F.,  2001,
  \mn@doi [\mnras] {10.1046/j.1365-8711.2001.04415.x}, \href
  {http://adsabs.harvard.edu/abs/2001MNRAS.325..257S} {325, 257}

\bibitem[\protect\citeauthoryear{{Storey}}{{Storey}}{1994}]{1994A&A...282..999S}
{Storey} P.~J.,  1994, \aap, \href
  {http://adsabs.harvard.edu/abs/1994A%26A...282..999S} {282, 999}

\bibitem[\protect\citeauthoryear{{Storey} \& {Hummer}}{{Storey} \&
  {Hummer}}{1995}]{1995MNRAS.272...41S}
{Storey} P.~J.,  {Hummer} D.~G.,  1995, \mn@doi [\mnras]
  {10.1093/mnras/272.1.41}, \href
  {http://adsabs.harvard.edu/abs/1995MNRAS.272...41S} {272, 41}

\bibitem[\protect\citeauthoryear{{Storey}, {Sochi}  \& {Badnell}}{{Storey}
  et~al.}{2014}]{2014MNRAS.441.3028S}
{Storey} P.~J.,  {Sochi} T.,   {Badnell} N.~R.,  2014, \mn@doi [\mnras]
  {10.1093/mnras/stu777}, \href
  {http://adsabs.harvard.edu/abs/2014MNRAS.441.3028S} {441, 3028}

\bibitem[\protect\citeauthoryear{{Tayal}}{{Tayal}}{2011}]{2011ApJS..195...12T}
{Tayal} S.~S.,  2011, \mn@doi [\apjs] {10.1088/0067-0049/195/2/12}, \href
  {http://adsabs.harvard.edu/abs/2011ApJS..195...12T} {195, 12}

\bibitem[\protect\citeauthoryear{{Tayal} \& {Gupta}}{{Tayal} \&
  {Gupta}}{1999}]{1999ApJ...526..544T}
{Tayal} S.~S.,  {Gupta} G.~P.,  1999, \mn@doi [\apj] {10.1086/307971}, \href
  {http://adsabs.harvard.edu/abs/1999ApJ...526..544T} {526, 544}

\bibitem[\protect\citeauthoryear{{Tayal} \& {Zatsarinny}}{{Tayal} \&
  {Zatsarinny}}{2010}]{2010ApJS..188...32T}
{Tayal} S.~S.,  {Zatsarinny} O.,  2010, \mn@doi [\apjs]
  {10.1088/0067-0049/188/1/32}, \href
  {http://adsabs.harvard.edu/abs/2010ApJS..188...32T} {188, 32}

\bibitem[\protect\citeauthoryear{{Toribio San Cipriano}, {Garc{\'{\i}}a-Rojas},
  {Esteban}, {Bresolin}  \& {Peimbert}}{{Toribio San Cipriano}
  et~al.}{2016}]{2016MNRAS.458.1866T}
{Toribio San Cipriano} L.,  {Garc{\'{\i}}a-Rojas} J.,  {Esteban} C.,
  {Bresolin} F.,   {Peimbert} M.,  2016, \mn@doi [\mnras]
  {10.1093/mnras/stw397}, \href
  {http://adsabs.harvard.edu/abs/2016MNRAS.458.1866T} {458, 1866}

\bibitem[\protect\citeauthoryear{{Tresse}, {Maddox}, {Loveday}  \&
  {Singleton}}{{Tresse} et~al.}{1999}]{1999MNRAS.310..262T}
{Tresse} L.,  {Maddox} S.,  {Loveday} J.,   {Singleton} C.,  1999, \mn@doi
  [\mnras] {10.1046/j.1365-8711.1999.02977.x}, \href
  {http://adsabs.harvard.edu/abs/1999MNRAS.310..262T} {310, 262}

\bibitem[\protect\citeauthoryear{{Trundle}, {Dufton}, {Lennon}, {Smartt}  \&
  {Urbaneja}}{{Trundle} et~al.}{2002}]{2002A&A...395..519T}
{Trundle} C.,  {Dufton} P.~L.,  {Lennon} D.~J.,  {Smartt} S.~J.,   {Urbaneja}
  M.~A.,  2002, \mn@doi [\aap] {10.1051/0004-6361:20021044}, \href
  {http://adsabs.harvard.edu/abs/2002A%26A...395..519T} {395, 519}

\bibitem[\protect\citeauthoryear{{Tsamis} \& {P{\'e}quignot}}{{Tsamis} \&
  {P{\'e}quignot}}{2005}]{2005MNRAS.364..687T}
{Tsamis} Y.~G.,  {P{\'e}quignot} D.,  2005, \mn@doi [\mnras]
  {10.1111/j.1365-2966.2005.09595.x}, \href
  {http://adsabs.harvard.edu/abs/2005MNRAS.364..687T} {364, 687}

\bibitem[\protect\citeauthoryear{{Tsamis}, {Barlow}, {Liu}, {Danziger}  \&
  {Storey}}{{Tsamis} et~al.}{2003}]{2003MNRAS.338..687T}
{Tsamis} Y.~G.,  {Barlow} M.~J.,  {Liu} X.-W.,  {Danziger} I.~J.,   {Storey}
  P.~J.,  2003, \mn@doi [\mnras] {10.1046/j.1365-8711.2003.06081.x}, \href
  {http://adsabs.harvard.edu/abs/2003MNRAS.338..687T} {338, 687}

\bibitem[\protect\citeauthoryear{{Urbaneja} et~al.,}{{Urbaneja}
  et~al.}{2005a}]{2005ApJ...622..862U}
{Urbaneja} M.~A.,  et~al., 2005a, \mn@doi [\apj] {10.1086/427468}, \href
  {http://adsabs.harvard.edu/abs/2005ApJ...622..862U} {622, 862}

\bibitem[\protect\citeauthoryear{{Urbaneja}, {Herrero}, {Kudritzki}, {Najarro},
  {Smartt}, {Puls}, {Lennon}  \& {Corral}}{{Urbaneja}
  et~al.}{2005b}]{2005ApJ...635..311U}
{Urbaneja} M.~A.,  {Herrero} A.,  {Kudritzki} R.-P.,  {Najarro} F.,  {Smartt}
  S.~J.,  {Puls} J.,  {Lennon} D.~J.,   {Corral} L.~J.,  2005b, \mn@doi [\apj]
  {10.1086/497528}, \href {http://adsabs.harvard.edu/abs/2005ApJ...635..311U}
  {635, 311}

\bibitem[\protect\citeauthoryear{{Venn}, {McCarthy}, {Lennon}, {Przybilla},
  {Kudritzki}  \& {Lemke}}{{Venn} et~al.}{2000}]{2000AAS...196.4011V}
{Venn} K.~A.,  {McCarthy} J.~K.,  {Lennon} D.~J.,  {Przybilla} N.,  {Kudritzki}
  R.~P.,   {Lemke} M.,  2000, in American Astronomical Society Meeting
  Abstracts \#196. p.~739

\bibitem[\protect\citeauthoryear{{Venn} et~al.,}{{Venn}
  et~al.}{2001}]{2001ApJ...547..765V}
{Venn} K.~A.,  et~al., 2001, \mn@doi [\apj] {10.1086/318424}, \href
  {http://adsabs.harvard.edu/abs/2001ApJ...547..765V} {547, 765}

\bibitem[\protect\citeauthoryear{{Zurita} \& {Bresolin}}{{Zurita} \&
  {Bresolin}}{2012}]{2012MNRAS.427.1463Z}
{Zurita} A.,  {Bresolin} F.,  2012, \mn@doi [\mnras]
  {10.1111/j.1365-2966.2012.22075.x}, \href
  {http://adsabs.harvard.edu/abs/2012MNRAS.427.1463Z} {427, 1463}

\bibitem[\protect\citeauthoryear{{de Vaucouleurs}, {de Vaucouleurs}, {Corwin},
  {Buta}, {Paturel}  \& {Fouque}}{{de Vaucouleurs}
  et~al.}{1991}]{1991S&T....82Q.621D}
{de Vaucouleurs} G.,  {de Vaucouleurs} A.,  {Corwin} Jr. H.~G.,  {Buta} R.~J.,
  {Paturel} G.,   {Fouque} P.,  1991, \skytel, \href
  {http://adsabs.harvard.edu/abs/1991S%26T....82Q.621D} {82, 621}

\makeatother
\end{thebibliography}



\appendix

\section{Line intensity ratios}\label{appex:1}

\begin{table*}
   \centering
   \caption{Line identifications and reddening corrected line fluxes with respect to \hb\ = 100 for the \hii\ regions 30 Doradus, N44C, IC2111, NGC1714 and N11B in the LMC.}
   \label{tab:lines_lmc}
   \begin{tabular}{lcccccccc}
        \hline
             $\lambda_0$     &       &      & $f(\lambda)$ &30 Doradus                 &    N44C                   &  IC2111                 & NGC1714                & N11B           \\
             \AA\            & Ion   &  ID  &              &$I$($\lambda$)             & $I$($\lambda$)            &$I$($\lambda$)           & $I$($\lambda$)         &$I$($\lambda$)  \\
        \hline                                                                                                                                                      
            3711.97          &\hi    &  H15 & 0.3824       & $1.63 \pm 0.05$           & $1.70 \pm 0.16$           &  $1.8 \pm 0.2$          &  $1.7 \pm 0.1$         & $1.7 \pm 0.1$ \\            
            3726.03          &\foii  &  1F  & 0.3762       & $ 63 \pm 1$               & $ 30 \pm 2$               & $ 128 \pm 6$            & $62 \pm 3$             & $102 \pm 8$     \\
            3728.82          &\foii  &  1F  & 0.3748       & $ 68 \pm 1$               & $ 39 \pm 3$               & $ 149 \pm 7$            & $66 \pm 3$             & $122 \pm 9$    \\
            3734.37          &\hi    &  H13 & 0.3725       & $2.26 \pm 0.06$           & $2.6 \pm 0.3$             &  $2.7 \pm 0.2$          & $1.7 \pm 0.1$          & $2.5 \pm 0.2$ \\
            3750.15          &\hi    &  H12 & 0.3655       & $2.95 \pm 0.07$           & $3.3 \pm 0.3$             &  $2.6 \pm 0.2$          & $3.4 \pm 0.2$          & $3.1 \pm 0.2$ \\
            3770.63          &\hi    &  H11 & 0.3566       & $3.10 \pm 0.06$           & $4.1 \pm 0.3$             &  $3.4 \pm 0.3$          & $4.3 \pm 0.2$          & $4.0 \pm 0.3$ \\
            3835.39          &\hi    &  H9  & 0.3291       & $7.7 \pm 0.1$             & $6.8 \pm 0.5$             &  $7.7 \pm 0.4$          & $7.3 \pm 0.4$          & $6.9 \pm 0.5$ \\
            3970.07          &\hi    &H$_\epsilon$&0.2751  & $15.7 \pm 0.2$            & $16 \pm 1.$               &  $16 \pm 1$             & $15.4 \pm 0.7$         & $16 \pm 1$  \\
            4068.6           &\fsii  &  1F  & 0.2384       & $ 0.86 \pm 0.03$          & $ 0.73 \pm 0.06$          & $ 1.2 \pm 0.1$          & $0.69 \pm 0.04$        & $1.2 \pm 0.1$ \\
            4076.35          &\fsii  &  1F  & 0.2354       & $ 0.29 \pm 0.02$          & $ 0.24 \pm 0.03$          & $ 0.42 \pm 0.03$        & $0.26 \pm 0.02$        & $0.42 \pm 0.04$ \\
            4101.74          &\hi    &H$_\delta$& 0.2262   & $26.9 \pm 0.3$            & $25 \pm 1$                & $24 \pm 1$              &  $25 \pm 1$            & $24 \pm 1$     \\
            4267.15          &\cii   &  6   & 0.1693       & $ 0.09 \pm 0.01$          & $ 0.127 \pm 0.005$        & $ 0.10 \pm 0.01$        & $0.11 \pm 0.01$        & $0.093 \pm 0.003$ \\
            4340.47          &\hi    &H$_\gamma$& 0.1457   & $ 46.8 \pm 0.3$           & $ 46 \pm 2$               & $ 46 \pm 1$             &  $ 45 \pm 2$           & $46 \pm 2 $    \\
            4363.21          &\foiii &  1F  & 0.1385       & $ 3.21 \pm 0.05$          & $ 6.8 \pm 0.5$            & $ 1.51 \pm 0.07$        & $2.5  \pm 0.1$         & $1.48 \pm 0.06$ \\
            4471.09          &\hei   &  14  & 0.1055       &        --                 & $3.44 \pm 0.14$           &          --             &       --               &      --           \\
            4638.86          &\oii   &  1   & 0.0578       & $ 0.049 \pm 0.005$        & $ 0.073 \pm 0.002$        & $ 0.040 \pm 0.009$      & $0.055 \pm 0.010$      & $0.048 \pm 0.002$ \\
            4641.81          &\oii   &  1   & 0.0570       & $ 0.063 \pm 0.006$        & $ 0.064 \pm 0.002$        & $ 0.054 \pm 0.010$      & $0.070 \pm 0.011$      & $0.045 \pm 0.002$ \\
            4649.13          &\oii   &  1   & 0.0550       & $ 0.053 \pm 0.005$        & $ 0.041 \pm 0.002$        & $ 0.050 \pm 0.009$      & $0.060 \pm 0.010$      & $0.049 \pm 0.002$ \\
            4650.84          &\oii   &  1   & 0.0545       & $ 0.061 \pm 0.006$        & $ 0.091 \pm 0.003$        & $ 0.049 \pm 0.009$      & $0.071 \pm 0.010$      & $0.061 \pm 0.002$  \\
            4661.63          &\oii   &  1   & 0.0516       & $ 0.071 \pm 0.007$        & $ 0.090 \pm 0.004$        & $ 0.045 \pm 0.009$      & $0.064 \pm 0.010$      & $0.057 \pm 0.002$  \\
            4673.73          &\oii   &  1   & 0.0484       & $ 0.010 \pm 0.003$        &      --                   & $ 0.014 \pm 0.006$      & $0.014 \pm 0.006$      & $0.011 \pm 0.001$  \\
            4676.24          &\oii   &  1   & 0.0477       & $ 0.021 \pm 0.003$        & $ 0.035 \pm 0.002$        & $ 0.022 \pm 0.007$      & $0.024 \pm 0.007$      & $0.018 \pm 0.001$  \\
            4685.68          &\heii  & 3.4  & 0.0452       &      --                   & $15.2 \pm  0.6$           &           --            &     --                 &      --            \\
            4696.36          &\oii   &  1   & 0.0423       & $ 0.003 \pm 0.001$        &    --                     & --                      &     --                 &      --            \\
            4861.33          &\hi    &\hb   & 0.0000       & $ 100 \pm  1$             & $100 \pm  3$              & $100 \pm  4$            &  $100 \pm  4$          & $100 \pm 3$          \\
            5006.84          &\foiii &  1F  &-0.0348       & $ 507 \pm 3$              & $ 704 \pm 11$             & $ 310 \pm 12$           & $ 444 \pm 18$          & $308 \pm 4$       \\
            5517.71          &\fcliii&  1F  &-0.1394       & $ 0.48 \pm 0.02$          & $ 0.33 \pm 0.02$          & $ 0.42 \pm 0.03$        & $ 0.42 \pm 0.02$       & $0.41 \pm 0.01$    \\
            5537.88          &\fcliii&  1F  &-0.1429       & $ 0.36 \pm 0.01$          & $ 0.25 \pm 0.01$          & $ 0.33 \pm 0.02$        & $ 0.33 \pm 0.02$       & $0.31 \pm 0.01$    \\
            5754.64          &\fnii  &  1F  &-0.1805       & $ 0.21 \pm 0.01$          & $ 0.10 \pm 0.01$          & $ 0.29 \pm 0.02$        & $ 0.15 \pm 0.02$       & $0.28 \pm 0.01$    \\
            5875.64          &\hei   &  11  &-0.2009       &          --               & $9.8 \pm 0.4$             &      --                 &       --               &         --         \\
            6312.1           &\fsiii &  3F  &-0.2709       & $ 2.15 \pm 0.04$          & $ 1.29 \pm 0.07$          & $ 1.56 \pm 0.08$        & $ 1.55 \pm 0.08$       & $1.46 \pm 0.07$    \\
            6562.82          &\hi    &H$_\alpha$&-0.3083   & $ 347 \pm  2$             & $ 284 \pm 18$             & $ 276 \pm 13$           &  $ 276 \pm 13$         & $292 \pm 17$        \\
            6583.41          &\fnii  &  3F  &-0.3114       & $ 14 \pm 1$               & $ 6.0 \pm 0.4$            & $ 21 \pm 1$             & $ 9.3 \pm 0.4$         & $20 \pm 1$         \\
            6678.15          &\hei   &  46  &-0.3248       &        --                 & $2.9 \pm 0.2$             &    --                   &       --               &        --          \\
            6716.47          &\fsii  &  2F  &-0.3300       & $ 8.8 \pm 0.1$            & $ 8.1 \pm 0.6$            & $ 13 \pm 1$             & $ 6.0  \pm 0.3$        & $13 \pm 1$         \\
            6730.85          &\fsii  &  2F  &-0.3322       & $ 8.1 \pm 0.1$            & $ 6.6 \pm 0.5$            & $ 11 \pm 1$             & $ 5.6 \pm 0.3$         & $11 \pm 1$         \\
            7318.92          &\foii  &  2F  &-0.4081       & $ 0.65 \pm 0.02$          & $ 0.27 \pm 0.03$          & $ 0.75 \pm 0.04$        & $ 0.53 \pm 0.03$       & $0.72 \pm 0.06$    \\
            7319.99          &\foii  &  2F  &-0.4082       & $ 1.97 \pm 0.05$          & $ 0.72 \pm 0.11$          & $ 2.2 \pm 0.1$          & $ 1.23 \pm 0.07$       & $2.1 \pm 0.2$    \\
            7329.67          &\foii  &  2F  &-0.4094       & $ 1.07 \pm 0.03$          & $ 0.44 \pm 0.04$          & $ 1.19 \pm 0.07$        & $ 0.80 \pm 0.05$       & $1.22 \pm 0.09$    \\
            7330.73          &\foii  &  2F  &-0.4095       & $ 1.06 \pm 0.03$          & $ 0.38 \pm 0.03$          & $ 1.20 \pm 0.07$        & $ 0.72 \pm 0.04$       & $1.16 \pm 0.09$    \\
            8392.40          &\hi    & P20  &-0.5208       &      --                   & $ 0.27 \pm 0.03$          &      --                 & $ 0.25 \pm 0.02$       & $0.26 \pm 0.03$  \\
            8413.32          &\hi    & P19  &-0.5227       & $ 0.38 \pm 0.02$          & $0.28 \pm 0.03$           & $ 0.29 \pm 0.02$        & $ 0.27 \pm 0.02$       & $0.27 \pm 0.03$  \\
            8437.96          &\hi    & P18  &-0.5249       & $ 0.35 \pm 0.02$          & $0.40 \pm 0.04$           & $ 0.32 \pm 0.02$        &      --                & $0.38 \pm 0.04$  \\
            8467.25          &\hi    & P17  &-0.5275       & $ 0.54 \pm 0.03$          & $0.39 \pm 0.04$           & $ 0.41 \pm 0.03$        & $ 0.41 \pm 0.03$       & $0.34 \pm 0.03$  \\
            8502.48          &\hi    & P16  &-0.5307       & $ 0.63 \pm 0.03$          & $0.46 \pm 0.05$           & $ 0.44 \pm 0.03$        & $ 0.47 \pm 0.03$       & $0.44 \pm 0.04$  \\
            8665.02          &\hi    & P13  &-0.5448       & $ 1.09 \pm 0.04$          & $0.79 \pm 0.09$           & $ 0.80 \pm 0.06$        & $ 0.83 \pm 0.05$       & $0.74 \pm 0.07$    \\
            8750.48          &\hi    & P12  &-0.5520       & $ 1.50 \pm 0.06$          &      --                   & $ 1.01 \pm 0.07$        & $ 1.22 \pm 0.08$       &      --           \\
            8862.79          &\hi    & P11  &-0.5612       & $ 1.89 \pm 0.06$          & $1.38 \pm 0.016$          & $ 1.31 \pm 0.09$        & $ 1.36 \pm 0.08$       & $1.3 \pm 0.1$    \\
            9014.91          &\hi    & P10  &-0.5732       &     --                    & $1.72 \pm 0.21$           & $ 1.3 \pm 0.1$          & $ 1.7 \pm 0.1$         & $1.7 \pm 0.2$    \\
            9068.9           &\fsiii &  1F  &-0.5774       & $ 42 \pm 1$               & $ 14 \pm 2$               & $ 42 \pm 3$             & $ 41 \pm 2$            & $23 \pm 2$          \\
            9229.01          &\hi    & P9   &-0.5894       & $3.8 \pm 0.1$             & $2.2 \pm 0.3$             & $ 2.4 \pm 0.2$          & $ 2.1 \pm 0.1$         & $2.2 \pm 0.2$     \\
            9545.97          &\hi    & P8   &-0.6119       &      --                   & $3.0 \pm 0.4$             &  --                     &  $ 2.7 \pm 0.2$        & $2.9 \pm 0.3$ \\
            10049.40         &\hi    & P7   &-0.6442       & $6.79 \pm 0.20$           & $6.3 \pm 1.8$             & $5.9 \pm 0.5$           &  $ 5.7 \pm 0.4$        & $6.1 \pm 0.7 $      \\
            $c$(\hb)         &       &      &              & $0.52 \pm 0.09$           & $0.21 \pm 0.09$           & $0.32 \pm 0.3$          & $0.31 \pm 0.2$         & $0.28 \pm 0.08$     \\
            \multicolumn{4}{l}{F(\hb)($\mathrm{erg\ cm^{-2}\ s^{-1}}$)}&               & $8.26 \times 10^{-13}$    & $1.41 \times 10^{-12}$  & $1.44 \times 10^{-12}$ & $7.81 \times 10^{-13}$ \\
          \hline
   \end{tabular}
\end{table*}

\begin{table*}
   \centering
   \caption{Line identifications and reddening corrected line fluxes with respect to \hb\ = 100 for the \hii\ regions N66, N81, NGC456 and N88A in the SMC.}
   \label{tab:lines_smc}
   \begin{tabular}{lccccccc}
        \hline
             $\lambda_0$     &       &      &              &N66A               &    N81                 &  NGC456              & N88A                \\
             \AA\            & Ion   &  ID  & $f(\lambda)$ &$I$($\lambda$)     & $I$($\lambda$)         &$I$($\lambda$)        & $I$($\lambda$)      \\
        \hline                                                                                                                                                                                                                                                                                                                                                                                                   
            3711.97          &\hi    &  H15 & 0.3824       & $1.8 \pm 0.1$     &  $1.8 \pm 0.1$         &   $1.4 \pm 0.1$      &  $2.1 \pm 0.3$   \\    
            3726.03          &\foii  &  1F  & 0.3762       & $77 \pm 5$        &    $51 \pm 3$          &   $68 \pm 1$         &   $27 \pm 4$        \\
            3728.82          &\foii  &  1F  & 0.3748       & $99 \pm 7$        &    $55 \pm 4$          &   $86 \pm 1$         &   $17 \pm 2$        \\
            3734.37          &\hi    &  H13 & 0.3725       & $2.6 \pm  0.2$    &  $2.7 \pm 0.2$         &   $1.9 \pm 0.1$      &  $2.9 \pm 0.4$   \\
            3750.15          &\hi    &  H12 & 0.3655       & $3.2 \pm 0.2$     &  $3.5 \pm 0.2$         &   $2.9 \pm 0.1$      &   $3.6 \pm 0.5$  \\            
            3770.63          &\hi    &  H11 & 0.3566       & $4.0 \pm 0.3$     &  $4.3 \pm 0.3$         &   $3.7 \pm 0.1$      &   $4.5 \pm 0.6$  \\            
            3835.39          &\hi    &  H9  & 0.3291       & $6.8 \pm 0.4$     &  $7.1 \pm 0.5$         &   $6.8 \pm 0.2$      &   $7.6 \pm 0.9$    \\            
            3970.07          &\hi    &H$_\epsilon$&0.2751  & $15.8 \pm 0.9$    &  $16.6 \pm 0.9$        &    --                &   $17 \pm 2$   \\            
            4068.6           &\fsii  &  1F  & 0.2384       & $1.2 \pm 0.1$     &    $0.60 \pm 0.04$     &   $1.02 \pm 0.08$    &   $0.75 \pm 0.06$   \\
            4076.35          &\fsii  &  1F  & 0.2354       & $0.42 \pm 0.04$   &    $0.20 \pm 0.02$     &   $0.33 \pm 0.04$    &   $0.25 \pm 0.02$   \\
            4101.74          &\hi    &H$_\delta$& 0.2262   & $25 \pm 1$        &  $25 \pm 1$            &   $26.0 \pm 0.2$     &   $27 \pm 2$    \\
            4267.15          &\cii   &  6   & 0.1693       & $0.035 \pm 0.004$ &    $0.038 \pm 0.002$   &   $0.04 \pm 0.02$    &   $0.042 \pm 0.002$  \\
            4340.47          &\hi    &H$_\gamma$& 0.1457   & $45 \pm 2$        &  $46 \pm 2$            &   $47.1 \pm 0.5$     &   $49 \pm 3$            \\             
            4363.21          &\foiii &  1F  & 0.1385 1     & $5.1 \pm 0.2$     &    $6.8 \pm 0.3$       &   $4.6 \pm 0.2$      &   $13 \pm 1$         \\
            4638.86          &\oii   &  1   & 0.0578       & --                &    $0.033 \pm 0.001$   &   $0.061 \pm 0.018$  &   $0.023 \pm 0.001$ \\
            4641.81          &\oii   &  1   & 0.0570       & --                &    $0.042 \pm 0.002$   &        *             &   $0.043 \pm 0.002$ \\
            4649.13          &\oii   &  1   & 0.0550       & $0.029 \pm 0.002$ &    $0.038 \pm 0.001$   &   $0.071 \pm 0.020$  &   $0.059 \pm 0.002$ \\
            4650.84          &\oii   &  1   & 0.0545       & $0.036 \pm 0.003$ &    $0.034 \pm 0.001$   &        *             &   $0.017 \pm 0.001$ \\
            4661.63          &\oii   &  1   & 0.0516       & $0.039 \pm 0.003$ &    $0.050 \pm 0.002$   &        --            &   $0.034 \pm 0.001$ \\
            4673.73          &\oii   &  1   & 0.0484       & --                &       --               &        --            &      --             \\
            4676.24          &\oii   &  1   & 0.0477       & --                &    $0.015 \pm 0.001$   &        --            &   $0.013 \pm 0.001$ \\
            4861.33          &\hi    &\hb   & 0.0000       & $100 \pm 3$       & $100 \pm 3$            &   $100 \pm 1$        &   $100 \pm 5$   \\
            5006.84          &\foiii &  1F  &-0.0348       & $403 \pm 6$       &    $503 \pm 8$         &   $396 \pm 4$        &   $671 \pm 13$       \\
            5517.71          &\fcliii&  1F  &-0.1394       & $0.34 \pm 0.01$   &    $0.37 \pm 0.01$     &   $0.41 \pm 0.04$    &   $0.25 \pm 0.01$    \\
            5537.88          &\fcliii&  1F  &-0.1429       & $0.25 \pm 0.01$   &    $0.28 \pm 0.01$     &   $0.29 \pm 0.04$    &   $0.32 \pm 0.02$    \\                        
            5754.64          &\fnii  &  1F  &-0.1805       & $0.16 \pm 0.01$   &    $0.08 \pm 0.1$      &   $0.14 \pm 0.03$    &   $0.07 \pm 0.01$    \\
            6312.1           &\fsiii &  3F  &-0.2709       & $1.79 \pm 0.08$   &    $1.76 \pm 0.08$     &   $1.82 \pm 0.09$    &   $1.8 \pm 0.2$    \\
            6562.82          &\hi    &H$_\alpha$&-0.3083   & $282 \pm 14$      &    $272 \pm 14$        &   $293 \pm 8$        &   $269 \pm 31$     \\
            6583.41          &\fnii  &  3F  &-0.3114       & $6.9 \pm 0.3$     &    $3.7 \pm 0.2$       &   $7.2 \pm 0.2$      &   $2.4 \pm 0.3$    \\
            6716.47          &\fsii  &  2F  &-0.3300       & $9.4 \pm 0.5$     &    $4.3 \pm 0.2$       &   $8.8 \pm 0.2$      &   $1.7 \pm 0.2$    \\
            6730.85          &\fsii  &  2F  &-0.3322       & $7.6 \pm 0.4$     &    $3.8 \pm 0.2$       &   $7.3 \pm 0.2$      &   $2.5 \pm 0.3$    \\
            7318.92          &\foii  &  2F  &-0.4081       & $0.73 \pm 0.05$   &    $0.50 \pm 0.03$     &   $3.2 \pm 0.1$      &   $0.60 \pm 0.09$    \\
            7319.99          &\foii  &  2F  &-0.4082       & $2.2 \pm 0.1$     &    $1.41 \pm 0.09$     &      *               &   $1.8 \pm 0.3$    \\
            7329.67          &\foii  &  2F  &-0.4094       & $1.2 \pm 0.1$     &    $0.78 \pm 0.05$     &   $2.6 \pm 0.1$      &   $0.94 \pm 0.1$    \\
            7330.73          &\foii  &  2F  &-0.4095       & $1.17 \pm 0.08$   &    $0.74 \pm 0.05$     &      *               &   $0.96 \pm 0.1$    \\            
            8392.40          &\hi    & P20  &-0.5208       & $0.26 \pm 0.02$   &   $0.24 \pm 0.02$      &   --                 &   $0.30 \pm 0.06$  \\
            8413.32          &\hi    & P19  &-0.5227       & $0.27 \pm 0.02$   &   $0.28 \pm 0.02$      & $0.18 \pm 0.03$      &   $0.46 \pm 0.09$  \\ 
            8437.96          &\hi    & P18  &-0.5249       & $0.34 \pm 0.03$   &   $0.33 \pm 0.03$      &      --              &   $0.39 \pm 0.07$  \\ 
            8467.25          &\hi    & P17  &-0.5275       & $0.38 \pm 0.03$   &   $0.38 \pm 0.03$      & $0.29 \pm 0.03$      &   $0.46 \pm 0.09$  \\ 
            8502.48          &\hi    & P16  &-0.5307       & $0.44 \pm 0.03 $  &   $0.45 \pm 0.04 $     &       --             &    $0.5 \pm 0.1$   \\ 
            8665.02          &\hi    & P13  &-0.5448       &     --            &   $1.03 \pm 0.09$      & $0.65 \pm 0.05$      &    $0.9 \pm 0.2$  \\ 
            8750.47          &\hi    & P12  &-0.5520       & $1.01 \pm 0.09$   &   $1.02 \pm 0.09$      &       --             &   $1.2 \pm 0.2$    \\
            8862.79          &\hi    & P11  &-0.5612       &      --           &   $1.8 \pm 0.2$        &       --             &   $1.6 \pm 0.3$    \\ 
            9014.91          &\hi    & P10  &-0.5732       & $1.7 \pm 0.2$     &   $1.7 \pm 0.1$        &       --             &   $2.0 \pm 0.4$    \\ 
            9068.9           &\fsiii &  1F  &-0.5774       & $23 \pm 2$        &    $14 \pm 1$          &      --              &   $14 \pm 3$         \\
            9229.01          &\hi    & P9   &-0.5894       & $2.2 \pm 0.2$     &   $2.2 \pm 0.2$        &       --             &   $2.6 \pm 0.5$    \\ 
            9545.97          &\hi    & P8   &-0.6119       & $2.7 \pm 0.3$     &   $2.8 \pm 0.2$        &       --             &   $3.6 \pm 0.8$      \\ 
            10049.40         &\hi    & P7   &-0.6442       & $5.7 \pm 0.6$     &   $5.7 \pm 0.6 $       &       --             &   $6.9 \pm 1.6$         \\             
            $c$(\hb)         &       &      &              & $0.21 \pm 0.07$   &    $0.11 \pm 0.07$     &   $0.17 \pm 0.03$    &   $0.48 \pm 0.16$   \\
            \multicolumn{4}{l}{F(\hb)($\mathrm{erg\ cm^{-2}\ s^{-1}}$)}&$4.88 \times 10^{-13}$& $2.47 \times 10^{-12}$&        &$3.45 \times 10^{-12}$ \\

            \hline
   \end{tabular}
\end{table*}


\bsp	
\label{lastpage}
\end{document}